\documentclass[twoside,twocolumn,9pt]{article}
\usepackage{extsizes}
\usepackage[super,sort&compress,comma]{natbib}
\usepackage[version=3]{mhchem}
\usepackage[left=1.5cm, right=1.5cm, top=1.785cm, bottom=2.0cm]{geometry}
\usepackage{balance}
\usepackage{mathptmx}
\usepackage{sectsty}
\usepackage{graphicx}
\usepackage{lastpage}
\usepackage[format=plain,justification=justified,singlelinecheck=false,font={stretch=1.125,small,sf},labelfont=bf,labelsep=space]{caption}
\usepackage{float}
\usepackage{fancyhdr}
\usepackage{fnpos}
\usepackage[english]{babel}
\addto{\captionsenglish}{%
  
}
\usepackage{array}
\usepackage{droidsans}
\usepackage{charter}
\usepackage[T1]{fontenc}
\usepackage[usenames,dvipsnames]{xcolor}
\usepackage{setspace}
\usepackage[compact]{titlesec}
\usepackage{hyperref}

\usepackage{mathtools}
\usepackage{physics}
\usepackage{siunitx}
	\DeclareSIUnit\molecule{molecule}
	\DeclareSIUnit\debye{D}
	\DeclareSIUnit\au{a.u.}
	\DeclareSIUnit\Buckingham{B}
        \DeclareSIUnit\bohr{\text{\ensuremath{a_0}}}
        \DeclareSIUnit\elementarycharge{\text{\ensuremath{e}}}
        \DeclareSIUnit\angstrom{\text{\AA}}
\usepackage{adjustbox}

\usepackage{epstopdf}

\definecolor{cream}{RGB}{222,217,201}

\newcommand{\ai}{\textit{ab initio}}

\newcommand{\Duo}{{\sc Duo}}

\newcommand{\oxygen}{\textsuperscript{16}O\textsubscript{2}}
\newcommand{\Xstate}{\ensuremath{\textrm{X}^3\,\Sigma^-_g}}
\newcommand{\astate}{\ensuremath{\textrm{a}^1\,\Delta_g}}
\newcommand{\bstate}{\ensuremath{\textrm{b}^1\,\Sigma^+_g}}
\newcommand{\Isstate}{\ensuremath{\textrm{I}^1\,\Pi_g}}
\newcommand{\Itstate}{\ensuremath{\textrm{I}^3\,\Pi_g}}
\newcommand{\IIsstate}{\ensuremath{\textrm{II}^1\,\Pi_g}}
\newcommand{\IItstate}{\ensuremath{\textrm{II}^3\,\Pi_g}}
\newcommand{\sstates}{\ensuremath{^1\,\Pi_g}}
\newcommand{\tstates}{\ensuremath{^3\,\Pi_g}}

\newcommand{\Mel}[3]{$\langle$#1$|#2|$#3$\rangle$}

\title{\bf An ab initio spectroscopic model of the molecular oxygen atmospheric and infrared
bands}

\author{Wilfrid Somogyi, Sergey N. Yurchenko,\footnote{s.yurchenko@ucl.ac.uk} \\
{\small \it Department of Physics and Astronomy, University College London, Gower Street, WC1E 6BT London} \\
\mbox{}\\
Gap-Sue Kim \\
{\small \it Dharma College, Dongguk University, 30, Pildong-ro 1-gil, Jung-gu, Seoul 04620, Korea}
}

\begin{document}

\maketitle

\begin{abstract}
We present a unified variational treatment of the magnetic dipole matrix elements, Einstein coefficients and line strength for general open-shell diatomic molecules in the general purpose diatomic code \Duo. Building on previous work in which similar expressions for the electric quadrupole transitions were developed, we also present a complete \ai\ spectroscopic model for the infrared, electric dipole-forbidden, spectrum of the \oxygen\ molecule. The model covers seven states, namely the \Xstate, \astate, \bstate, \Isstate, \IIsstate, \Itstate and \IItstate\ states, for which 7 potential energy, 6 electronic angular momentum, 7 spin-orbit, and 14 quadrupole moment curves are calculated using ic-MRCI theory and an aug-cc-pV5Z basis set. These curves are diabatised to remove avoided crossings between the excited $\Pi$ states, and the resultant properties are used to produce a line list for higher-order transitions of astrophysical interest.
\end{abstract}



\begin{figure}[h]
    \centering
    \includegraphics[width=\linewidth]{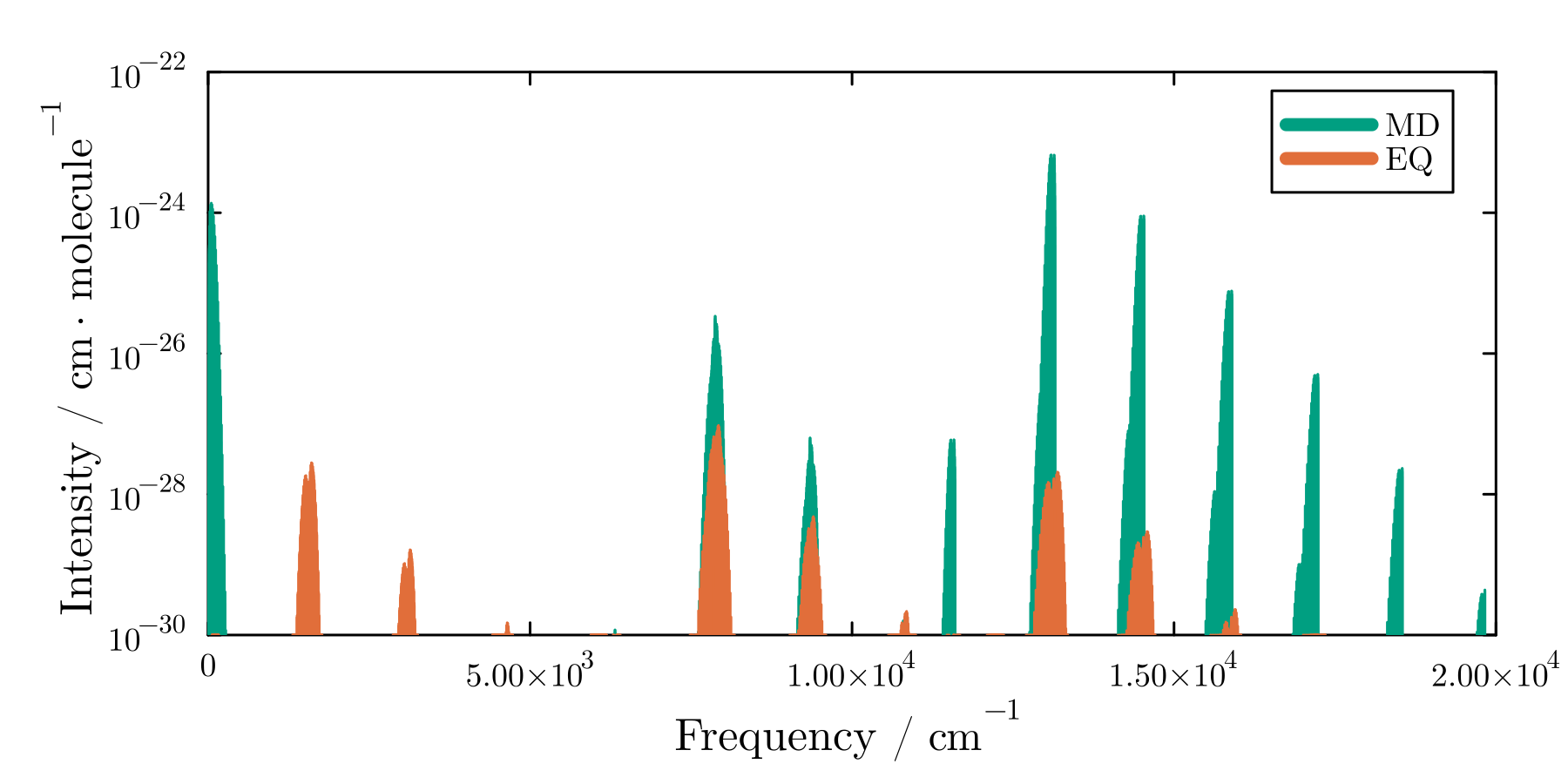}
    \caption{The electric quadrupole and magnetic dipole line list at \SI{296}{\kelvin} for \oxygen\ in the wavenumber range \SIrange{0}{18000}{\per\cm}.}
    \label{fig:abinitio_line_list}
\end{figure}

\section{Introduction}
\label{sec:introduction}

Oxygen is the most abundant element on Earth, and the third most abundant element in the universe. In its diatomic molecular form, \oxygen, it is also the second most abundant molecule in Earth's atmosphere and consequently plays a critical role in the metabolism of living organisms. Aside from its biological importance, oxygen also plays an important part in many chemical reactions, both industrial and geophysical. Due to it's ubiquity, \oxygen\ has been the focus of many studies and reviews spanning more than seven decades \citep{62Waxxxx.O2, 72Krxxxx.O2, 79HuHexx.O2, 88SlCoxx.O2, 12DrGuYu.O2, 12YuMiDr.O2, 13DrYuEl.O2, 14YuDrMi.O2}. The \oxygen\ molecule is also of special interest in an exoplanetary context, owing to both it's geophysical role\cite{19DoYoKl.O2, 21LiBoHa.O2, 21LivaMa.O2, 16ScWoBe.O2, 17ScFexx.O2, 20KoMaZo.O2} and its potential as a biosignature on Earth-like exoplanets\cite{18MeReAr.O2, 21KrFoNi.O2, 20LiDeUn.O2, 22KrThGa.O2}.

Detecting the presence of \oxygen\ in exoplanetary atmospheres requires spectral characterisation across a broad range of frequencies, particularly in the infra-red region of the electromagnetic spectrum, where telescopes such as JWST and the ELT are primed to perform measurements of astronomical spectra\citep{16GrLiMo.exo, 18DaBaLa.exo, 22KrThGa.O2, 19LoBeGo.O2, 23HaApBe.O2}. The existing body of work has produced extremely accurate spectrosopic data for a number of transitions, focused primarily on the narrow region of the spectrum that corresponds to Earth's atmospheric bands. However,  Due to the limitations of experimental techniques, a broader characterisation of the \oxygen\ spectrum has yet to be completed.

Owing to it's molecular symmetry, the homonuclear diatomic \oxygen\ molecule has zero permanent electric dipole moment (E1) and, as a result, pure vibrational and rotational transitions are forbidden in the electric dipole approximation. The three lowest lying electronic states ($\Xstate$, $\astate$ and $\bstate$), which comprise the infrared spectrum, all have \textit{gerade} symmetries. Thus transitions between them are also forbidden by electric dipole selection rules. The strongest absorption bands for the oxygen molcule are electric dipole bands in the ultraviolet region. These include transitions to lowest energy \textit{ungerade} states are $\mathrm{c}^1\Sigma_u^-$, $\mathrm{A}'^3\Delta_u$ and $\mathrm{A}^3\Sigma_u^+$, (known as the Herzberg bands) and the $\mathrm{B}^3\Sigma^-_u$ -- \bstate\ bands, which borrow intensity from the electric dipole Schumman-Runge $\mathrm{B}^3\Sigma^-_u$ -- \Xstate\ transitions \citep{54BrGaxx.O2, 84YoFrPa.O2, 99JeMeCo.O2, 99LeGiSl.O2, 00Mixxxx.O2}. However, the long wavelength of these absorption bands make them unlikely candidates for detection in an exoplanetary context. Telescopes for atmospheric remote sensing, such as CHEOPS, JWST, ARIEL and PLATO target the infrared and visible region, which is dominated in \oxygen\ by higher order magnetic dipole (M1) and electric quadrupole (E2) transitions.

The atmospheric $A$, $B$ and $\gamma$-bands, due to the fundamental and first two overtone $\bstate \leftarrow \Xstate$ transitions \citep{87RiWixx.O2, 99ScLexx.O2, 09LoHaOk.O2, 11GoRoTo.O2, 17DrBeBr.O2}, and infrared transitions in the ground electronic state have been the focus of numerous studies dating back to the 1920s \citep{1927DiBaxx.O2, 47HeHexx.O2, 52Hexxxx.O2, 53Hexxxx.O2, 87ZiMixx.O2}. The structure of these bands was explained primarily by M1 interactions \citep{50BuAnSm.O2, 48BaHexx.O2, 10DrYuMi.O2}, and although it was known that these should be accompanied by significantly weaker E2 lines \citep{48BaHexx.O2}, they were not observed in terrestrial spectra until 1980 \citep{80Brxxxx.O2}. Infrared E2 transitions within the ground electronic state were also observed in 1981 \citep{81GoReRo.O2, 81RoGoxx.O2} and more recently measured in the laboratory \citep{20ToBeOd.O2}. Additional quadrupole bands, due to the \astate\ -- \Xstate\ transition \citep{10LeKaGo.O2, 47HeHexx.O2}, have also been observed in solar spectra \citep{10GoKaCa.O2}, and the magnetic dipole counterparts in atmospheric afterglow \citep{81AmVexx.O2}.

The Noxon band \citep{61Noxxxx.O2}, which consists of transitions between the first two excited states, \bstate\ -- \astate, is also of interest. It is forbidden for both the E1 and M1 moments due to, respectively, the \textit{gerade} symmetry of both states and the $\Delta \Lambda = 2$ property. The absence of spin-orbit coupling between the two states means the Noxon band is purely electric quadrupolar in nature. Additionally, the same spin-orbit coupling between the two $\Sigma_g$ states that generates the \bstate\ -- $\mathrm{B}^3\Sigma^-_u$ transitions results in the Noxon band contributing significantly to the quadrupolar intensity of the \bstate\ -- \Xstate\ and \astate\ -- \Xstate\ bands, a phenomenon first described by \citet{83SvMixx} and subsequently observed in various experiments \citep{80Mixxxx.O2, 11MiBaSh.O2, 94BaNaxx.O2}.

The ExoMol project aims to provide accurate line lists for a wide range of molecules relevant to exoplanetary applications. In the case of diatomic molecules, the \Duo\ program allows one to construct spectroscopic models for diatomic molecules with an arbitrary number of potential energy curves and couplings. These can be represented as either a grid of \ai\ values or as an analytic representation using the various functional forms that are supported. For a full account of the \Duo\ code we refer the reader to \citet{16YuLoTe.methods}. Recent extensions to this program now allow for the calculation of magnetic dipole and electric quadrupole line strengths, which are crucial to the spectra of many homonuclear molecules such as \oxygen. Once the absorption intensities have been computed, the \textsc{ExoCross} program allows one to create absorption cross-sections for a range of line profiles that account for effects such as collisional broadening and temperature dependence \citep{ExoCross}. In the case of molecular oxygen, for example, the terrestrial absorption lines are strongly determined by collisions with N\textsubscript{2} and other O\textsubscript{2} molecules \citep{17KavaGr.O2}. However, for other exoplanets, these effects are likely to vary considerably, and such considerations must be made on a case-by-case basis depending on the context (atmospheric pressure, abundance, etc.) in which the molecule at hand is studied.

A considerable amount of effort has been devoted to establishing highly accurate potential energy curves for a range of electronic states of \oxygen. 
In 2010 \citet{10ByMaRu.O2} employed Dunning's correlation-consistent triple and quadruple-zeta basis sets to determine the full configuration interaction energy for the three lowest-lying electronic states; $\Xstate$, $\astate$, and $\bstate$, including core-valence correlations and relativistic contributions. Using a method developed in earlier work, which they term \textit{correlation energy extrapolation by intrinsic scaling}, they extrapolate these energies to the complete basis set limit  \citep{10ByMaRu.O2, 10ByRuxx.O2}. In 2014 \citet{14LiShSu.O2} produced accurate PECs for 22 electronic states as well 54 spin-orbit coupling curves via the complete active space self-consistent field (CASSCF) method, and a subsequent internally contracted multireference configuration interaction calculation including the Davidson correction (icMRCIQ).

\section{Magnetic Dipole Linestrengths}

In a previous work we reported expressions for the electric quadrupole linestrengths. These expressions were implemented in the variational spectroscopic program \Duo \cite{Duo}, and verified using accurate spectral data for a variety of molecules, including H\textsubscript{2} and the Noxon band of \oxygen. Here we report similar expressions for general magnetic dipole transitions in diatomic molecules.
\Duo\ uses the Hund's case (a) basis set in the following form:
\begin{equation}
    \ket{\varphi_i} =
    \ket{\xi \Lambda}\ket{S \Sigma}\ket{\xi v}\ket{J \Omega M}
\label{eq:duo_basis}
\end{equation}
where $J$ is the total angular momentum, $M$ is a projection of $J$ on the laboratory $Z$-axis in units of $\hbar$, $S$ is the total electronic spin angular momentum, $\Sigma$ is the projection of the spin of electrons on molecular $z$-axis, $\xi$ are indexes of the $\xi$-th electronic state, $\Lambda$ is the projection of the electronic angular momentum on molecular $z$-axis, $\Omega = \Lambda + \Sigma$ (projection of the total angular momentum on molecule $z$-axis) and $v$ is the vibrational quantum number.

The linestrengths in the representation of these basis set functions are then given by
\begin{multline}
    S(\mathrm{f} \gets \mathrm{i}) = g_\mathrm{ns} (2J_i + 1) (2J_f + 1) \left| \sum_{\varphi_\mathrm{f}} C_{J_\mathrm{f} \tau_\mathrm{f}}^* (\varphi_\mathrm{f}) \sum_{\varphi_\mathrm{i}} C_{J_\mathrm{i} \tau_\mathrm{i}} (\varphi_\mathrm{i}) \sum_{m'} (-1)^{m'+\Omega_\mathrm{i}} \vphantom{\begin{pmatrix} J \\ J \end{pmatrix}} \right. \\
    \left. \times \mel{S_f \Sigma_f}{
            \mel{v_f}{
                \mel{\varepsilon_f \Lambda_f}{d_{m'}}{\varepsilon_i \Lambda_i}
            }{v_i}
        }{S_i \Sigma_i}
    \begin{pmatrix}
        J_\mathrm{i} & J_\mathrm{f} & 1 \\
        -\Omega_\mathrm{i} & \Omega_\mathrm{f} & -m'
    \end{pmatrix}
    \right|^2
\end{multline}
where $m'$ indexes the spherical components of the magnetic dipole moment in the molecule-fixed frame. The Wigner $3j$ symbol implies the following selection rules
\begin{align}
    \Delta J &= J_f - J_i = 0, \pm 1, \\
    \Delta \Omega &= \Omega_f - \Omega_i = m'.
\end{align}

The magnetic dipole moment $\vec{d}$ arises due to the motion of charged particles. In the case of diatomic molecules, the primary magnetic dipole moment is due to negatively charged electrons with non-zero angular momenta. This magnetic moment is the sum of the orbital electronic angular momentum, and the spin electronic angular momentum as
\begin{equation}
    \vec{d} = \vec{L} + g_s \vec{S},
\end{equation}
where $g_s$ is the spin Lande g-factor. The matrix elements can then be written in terms of the raising and lowering operators $\hat{L}_{\pm}$ and $\hat{S}_{\pm}$
\begin{align}
    \begin{split}
        &\mel{S_f \Sigma_f}{
                \mel{v_f}{
                    \mel{\varepsilon_f \Lambda_f}{\hat{L}_{+1} + g_s \hat{S}_{+1}}{\varepsilon_i \Lambda_i}
                }{v_i}
            }{S_i \Sigma_i} \\
        & \qquad = -\frac{1}{\sqrt{2}}
        \left[
            \mel{v_f}{\hat{L}_+(r; \xi_i, \xi_f)}{v_i} \delta_{\Sigma_f \Sigma_i} \delta_{\Lambda_f \Lambda_i + 1} \right. \\
        & \qquad \qquad \left.
            +  g_s \left[ S_i(S_i + 1) - \Sigma_i(\Sigma_i + 1) \right]^\frac{1}{2} \delta_{v_f v_i} \delta_{\xi_f \xi_i} \delta_{\Lambda_f \Lambda_i} \delta_{\Sigma_f \Sigma_i + 1}
        \right] \delta_{S_f S_i},
    \end{split} \\
    \begin{split}
        &\mel{S_f \Sigma_f}{
                \mel{v_f}{
                    \mel{\varepsilon_f \Lambda_f}{\hat{L}_{-1} + g_s \hat{S}_{-1}}{\varepsilon_i \Lambda_i}
                }{v_i}
            }{S_i \Sigma_i} \\
        & \qquad = \frac{1}{\sqrt{2}}
        \left[
            \mel{v_f}{\hat{L}_-(r; \xi_i, \xi_f)}{v_i} \delta_{\Sigma_f \Sigma_i} \delta_{\Lambda_f \Lambda_i - 1}
        \right. \\
        & \qquad \qquad \left.
            +  g_s \left[ S_i(S_i + 1) - \Sigma_i(\Sigma_i - 1) \right]^\frac{1}{2} \delta_{v_f v_i} \delta_{\xi_f \xi_i} \delta_{\Lambda_f \Lambda_i} \delta_{\Sigma_f \Sigma_i - 1}
        \right] \delta_{S_f S_i},
    \end{split} \\
    \begin{split}
        & \mel{S_f \Sigma_f}{
                \mel{v_f}{
                    \mel{\varepsilon_f \Lambda_f}{\hat{L}_{0} + g_s \hat{S}_{0}}{\varepsilon_i \Lambda_i}
                }{v_i}
            }{S_i \Sigma_i} \\
        & \qquad = (\Lambda_i + g_s \Sigma_i) \delta_{v_f v_i} \delta_{\xi_f \xi_i} \delta_{\Lambda_f \Lambda_i} \delta_{S_f S_i} \delta_{\Sigma_f \Sigma_i}.
    \end{split}
\end{align}

\section{Intensity Structure of the Infrared Bands \bstate\ -- \Xstate, \astate\ -- \Xstate and \Xstate\ -- \Xstate}
\label{sec:structure_of_the_atmospheric_bands}

The composition of the quadrupole and magnetic intensities of the forbidden atmospheric bands of \oxygen\ has been expounded in detail throughout the literature\citep{78Mixxxx.O2, 80Mixxxx.O2, 94BaNaxx.O2, 11MiBaSh.O2, 13MiMuAg.O2}. In particular, we highlight the work of \citet{13MiMuAg.O2}, who provide a comprehensive account of individual contributions to the line strength of the \bstate\ -- \Xstate\, \astate\ -- \Xstate, and Noxon \bstate\ -- \astate\ bands. The electric quadrupole transitions are, in general, weaker than magnetic dipole transitions, but nonetheless have been observed in both the laboratory \citep{61Noxxxx.O2, 86FiKrRa.O2, 09LoHaOk.O2} and in atmospheric solar spectra \citep{10GoKaCa.O2, 80Brxxxx.O2}. They are present in both the \bstate\ -- \Xstate, and \astate\ -- \Xstate\ bands. Crucial to an account of the atmospheric bands are a set of highly excited $\Pi$ states, namely the $1^1\Pi_g$, $2^1\Pi_g$, $1^3\Pi_g$, and $2^3\Pi_g$ states \citep{86KlPexx.O2, 11MiBaSh.O2}. The two pairs of states with the same spin multiplicity each exhibit avoided crossings, with the $2^3\Pi_g$ and $2^1\Pi_g$ states being pre-dissociative in nature. In order to treat this collection of $\Pi$ states we transform to the diabatic representation\cite{22BrYuKi.SO} to produce the \IIsstate, \Isstate, \IItstate, and \Itstate\ states. The \IIsstate\ and \IItstate\ exhibit shallow potential wells with minima approximately \SI{60000}{\per\cm} above the zero-point energy, which give rise to a small number of bound vibrational levels. In the following, we analyse the main contributions to the electric quadrupole and magnetic dipole line strengths for three systems \bstate\ -- \Xstate\, \astate\ -- \Xstate and \Xstate\ -- \Xstate.

\subsection{Electric Quadrupole Transitions}

The \bstate\ -- \Xstate\ electric quadrupole intensities are comprised of two quadrupole moments corresponding to components of the irreducible quadrupole moment operator, (see \citet{21SoYuYa} for discussion of the irreducible representation). The first is the $Q^{(2)}_0$ moment, which arises as a result of spin-orbit mixing between the \Xstate\ and \bstate\ states, and generates two diagonal contributions to the total intensity for $\Delta \Omega = 0$ transitions. The second is the $Q^{(2)}_{\pm 1}$ component, which generates $\Delta \Omega = \pm 1$ transitions via spin-orbit coupling of the \Xstate\ state and the excited \sstates\ states. Here $\Omega = \Lambda + \Sigma$ is the projection of the total electronic angular momentum on the molecular axis, with $\Lambda$ and $\Sigma$ the orbital and spin angular momenta respectively.  Denoting the first-order SOC-perturbed expressions with a subscript $p$, for the \bstate\ -- \Xstate\ line strength we write
\begin{align}
    \begin{split}
        &\mel{\mathrm{b}^1\Sigma^+_g}{Q^{(2)}_0}{\mathrm{X}^3\Sigma^-_{g,0}}_p \\
        & \quad = \frac{\mel{\mathrm{b}^1\Sigma^+_g}{H_\text{SO}}{\mathrm{X}^3\Sigma^-_{g, 0}}}{E(\mathrm{X}^3\Sigma^-_{g, 0}) - E(\mathrm{b}^1\Sigma^+_g)} \mel{\mathrm{b}^1\Sigma^+_g}{Q^{(2)}_0}{\mathrm{b}^1\Sigma^+_g} \\
        &\qquad \qquad + \frac{\mel{\mathrm{X}^3\Sigma^-_{g, 0}}{H_\text{SO}}{\mathrm{b}^1\Sigma^+_g}^*}{E(\mathrm{b}^1\Sigma^+_g) - E(\mathrm{X}^3\Sigma^-_{g, 0}) } \mel{\mathrm{X}^3\Sigma^-_{g, 0}}{Q^{(2)}_0}{\mathrm{X}^3\Sigma^-_{g, 0}} \label{eq:b-X_Q20}
    \end{split} \\
    &\qquad = \alpha_0 \mel{\mathrm{b}^1\Sigma^+_g}{Q^{(2)}_0}{\mathrm{b}^1\Sigma^+_g} - \alpha_0\mel{\mathrm{X}^3\Sigma^-_{g, 0}}{Q^{(2)}_0}{\mathrm{X}^3\Sigma^-_{g, 0}} \\
    \begin{split}
        &\mel{\mathrm{b}^1\Sigma_g}{Q^{(2)}_{\pm 1}}{\mathrm{X}^3\Sigma^-_{g, 1}}_p = \\
        & \quad \sum_{\zeta} \frac{\mel{\zeta^1\Pi_g}{H_\text{SO}}{\mathrm{X}^3\Sigma^-_{g, \pm 1}}}{E(\mathrm{X}^3\Sigma^-_{g, 1}) - E(\zeta^1\Pi_g)} \mel{\mathrm{b}^1\Sigma^+_g}{Q^{(2)}_{\pm 1}}{\zeta^1\Pi_g}
    \end{split} \\
    & \qquad = \sum_{\zeta} \alpha_{\zeta, \mathrm{X}} \mel{\mathrm{b}^1\Sigma^+_g}{Q^{(2)}_{\pm 1}}{\zeta^1\Pi_g} \label{eq:b-X_Q21}.
\end{align}
The \sstates\ states lies far in energy above the \Xstate, and the thus primary contribution to the \bstate\ -- \Xstate\ band comes from the difference in permanent electric quadrupole moments of the two $\Sigma$ states.

We now turn to the quadrupole line strength of the \astate\ -- \Xstate\ band, which consists of two components, as given by
\begin{align}
    \begin{split}
        &\mel{\mathrm{a}^1\Delta_g}{Q^{(2)}_{\pm 2}}{\mathrm{X}^3\Sigma^-_{g,0}}_p \\
        & \quad =
        \frac{\mel{\mathrm{b}^1\Sigma^+_g}{H_\text{SO}}{\mathrm{X}^3\Sigma^-_{g, 0}}}{E(\mathrm{X}^3\Sigma^-_{g, 0}) - E(\mathrm{b}^1\Sigma^+_g)} \mel{\mathrm{a}^1\Delta_g}{Q^{(2)}_{\pm 2}}{\mathrm{b}^1\Sigma^+_g} \\
        &\qquad + \sum_{\xi} \frac{\mel{\xi^3\Pi_{g, 2}}{H_\text{SO}}{\mathrm{a}^1\Delta_g}^*}{E(\mathrm{a}^1\Delta_g) - E(\xi^3\Pi_{g, 2})} \mel{\xi^3\Pi_{g, 2}}{Q^{(2)}_{\pm 2}}{\mathrm{X}^3\Sigma^-_{g, 0}}
    \end{split} \\
        &\quad =\alpha_{\mathrm{b}, \mathrm{X}} \mel{\mathrm{a}^1\Delta_{g, 2}}{Q^{(2)}_{\pm2}}{\mathrm{b}^1\Sigma^+_{g, 0}}  + \sum_{\xi} \alpha_{\xi, \mathrm{a}}^* \mel{\xi^3\Pi_{g, 2}}{Q^{(2)}_{\pm 2}}{\mathrm{X}^3\Sigma^-_{g, 0}} \label{eq:a-X_Q22} \\
    \begin{split}
        &\mel{\mathrm{a}^1\Delta_g}{Q^{(2)}_{\pm1}}{\mathrm{X}^3\Sigma^-_{g,1}}_p \\
        &\quad =
        \sum_{\zeta} \frac{\mel{\zeta^1\Pi_g}{H_\text{SO}}{\mathrm{X}^3\Sigma^-_{g, 1}}}{E(\mathrm{X}^3\Sigma^-_{g, 1}) - E(\zeta^1\Pi_g)}
        \mel{\mathrm{a}^1\Delta_g}{Q^{(2)}_{\pm 1}}{\zeta^1\Pi_g} \\
        &\qquad + \sum_{\xi} \frac{\mel{\xi^3\Pi_{g, 2}}{H_\text{SO}}{\mathrm{a}^1\Delta_g}^*}{E(\mathrm{a}^1\Delta_g) - E(\xi^3\Pi_{g, 2})}
        \mel{\xi^3\Pi_{g, 2}}{Q^{(2)}_{\pm 1}}{\mathrm{X}^3\Sigma^-_{g, 1}}
    \end{split} \\
    &= \sum_\zeta \alpha_{\zeta, \mathrm{X}} \mel{\mathrm{a}^1\Delta_g}{Q^{(2)}_{\pm1}}{\zeta^1\Pi_g} + \sum_\xi \alpha_{\xi, \mathrm{a}}^* \mel{\xi^3\Pi_{g,2}}{Q^{(2)}_{\pm1}}{\mathrm{X}^3\Sigma^-_{g,1}}  \label{eq:a-X_Q21}
\end{align}
The first component arises primarily from $Q^{(2)}_{\pm 2}$, and borrows strength from the Noxon band (\bstate\ -- \astate) via spin-orbit mixing of the \Xstate\ with the \bstate\ state. There is an additional contribution to this band from the \tstates\ -- \Xstate\ quadrupole moments due to spin-orbit mixing of the \astate\ state and the \tstates\ states. The same \tstates\ -- \Xstate\ quadrupole moment also contributes to the second (weaker) component of the \astate\ -- \Xstate\ line strength, via $Q^{(2)}_{\pm1}$, together with the \astate\ -- \sstates\ quadrupole moment due to spin-orbit mixing between the \Xstate\ and \sstates\ states. The dominant contribution comes from the Noxon band, due to the fact that the \bstate\ state and \astate\ state are closely separated in energy, and the transition quadrupole moment between them is reasonably strong. This hypothesis is supported by the work of \citet{11MiBaSh.O2}, in which the ratio $Q^{(2)}_{\pm1} / Q^{(2)}_{\pm2}$ is calculated for various lines with a value in the range \numrange{0.012}{0.055}.

Finally, the electric quadrupole moment contribution to the \Xstate\ -- \Xstate\ intensities arise predominantly from the diagonal quadrupole moment of the ground electronic state, but with additional contributions from the spin-orbit coupling with the \sstates

\begin{align}
    \begin{split}
        &\mel{\mathrm{X}^3\Sigma^-_{g, 0}}{Q^{(2)}_0}{\mathrm{X}^3\Sigma^-_{g, 0}}_p \\
        &\quad = \mel{\mathrm{X}^3\Sigma^-_{g, 0}}{Q^{(2)}_0}{\mathrm{X}^3\Sigma^-_{g, 0}} \\
        &\qquad + \frac{\mel{\mathrm{X}^3\Sigma^-_{g, 0}}{H_\text{SO}}{\mathrm{b}^1\Sigma^+_g}}{E(\mathrm{b}^1\Sigma^+_g) - E(\mathrm{X}^3\Sigma^-_{g, 0})}\mel{\mathrm{b}^1\Sigma^+_g}{Q^{(2)}_0}{\mathrm{X}^3\Sigma^-_{g, 0}} \\
        &\qquad + \frac{\mel{\mathrm{b}^1\Sigma^+_g}{H_\text{SO}}{\mathrm{X}^3\Sigma^-_{g, 0}}}{E(\mathrm{X}^3\Sigma^-_{g, 0}) - E(\mathrm{b}^1\Sigma^+_g)}\mel{\mathrm{X}^3\Sigma^-_{g, 0}}{Q^{(2)}_0}{\mathrm{b}^1\Sigma^+_g}
    \end{split} \\
    \begin{split}
        &\quad = \mel{\mathrm{X}^3\Sigma^-_{g, 0}}{Q^{(2)}_0}{\mathrm{X}^3\Sigma^-_{g, 0}} - \alpha_{\mathrm{b}, \mathrm{X}}^* \mel{\mathrm{b}^1\Sigma^+_g}{Q^{(2)}_0}{\mathrm{X}^3\Sigma^-_{g, 0}} \\
        &\qquad + \alpha_{\mathrm{b}, \mathrm{X}} \mel{\mathrm{X}^3\Sigma^-_{g, 0}}{Q^{(2)}_0}{\mathrm{b}^1\Sigma^+_g}
        \label{eq:X-X_Q20_0}
    \end{split} \\
    \begin{split}
        &\mel{\mathrm{X}^3\Sigma^-_{g, 1}}{Q^{(2)}_0}{\mathrm{X}^3\Sigma^-_{g, 1}}_p \\
        &\quad = \mel{\mathrm{X}^3\Sigma^-_{g, 1}}{Q^{(2)}_0}{\mathrm{X}^3\Sigma^-_{g, 1}} \\
        &\qquad + \sum_\zeta \frac{\mel{\mathrm{X}^3\Sigma^-_{g, 1}}{H_\text{SO}}{\zeta^1\Pi_g}}{E(\zeta^1\Pi_g) - E(\mathrm{X}^3\Sigma^-_{g, 1})}\mel{\zeta^1\Pi_g}{Q^{(2)}_0}{\mathrm{X}^3\Sigma^-_{g, 1}} \\
        &\qquad + \frac{\mel{\zeta^1\Pi_g}{H_\text{SO}}{\mathrm{X}^3\Sigma^-_{g, 1}}}{E(\mathrm{X}^3\Sigma^-_{g, 1}) - E(\zeta^1\Pi_g)}\mel{\mathrm{X}^3\Sigma^-_{g, 1}}{Q^{(2)}_0}{\zeta^1\Pi_g}
    \end{split} \\
    \begin{split}
        &\quad = \mel{\mathrm{X}^3\Sigma^-_{g, 0}}{Q^{(2)}_0}{\mathrm{X}^3\Sigma^-_{g, 0}} - \sum_\zeta \alpha_{\zeta, \mathrm{X}}^* \mel{\zeta^1\Pi_g}{Q^{(2)}_0}{\mathrm{X}^3\Sigma^-_{g, 1}} \\
        &\qquad + \alpha_{\zeta, \mathrm{X}} \mel{\mathrm{X}^3\Sigma^-_{g, 1}}{Q^{(2)}_0}{\zeta^1\Pi_g} \label{eq:X-X_Q20_1}
    \end{split}
\end{align}

\subsection{Magnetic Dipole Transitions}

The strongest of the \oxygen\ atmospheric transitions are the magnetic dipole transitions in the \bstate -- \Xstate\ band  with the main contributions to the line strengths given by
\begin{align}
    \begin{split}
        &\mel{\mathrm{b}^1\Sigma^+_g}{d_{\pm 1}}{\mathrm{X}^3\Sigma^-_{g, 1}}_p \\
        &\quad = \frac{\mel{\mathrm{X}^3\Sigma^-_{g, 0}}{H_\text{SO}}{\mathrm{b}^1\Sigma^+_g}^*}{E(\mathrm{b}^1\Sigma^+_g) - E(\mathrm{X}^3\Sigma^-_{g, 0})} \mel{\mathrm{X}^3\Sigma^-_{g, 0}}{\hat{S}_{\pm 1}}{\mathrm{X}^3\Sigma^-_{g, 1}} \\
        &\qquad + \sum_{\zeta} \frac{\mel{\zeta^1\Pi_g}{H_\text{SO}}{\mathrm{X}^3\Sigma^-_{g, 1}}}{E(\mathrm{X}^3\Sigma^-_{g, 1}) - E(\zeta^1\Pi_g)} \mel{\mathrm{b}^1\Sigma^+_g}{\hat{L}_{\pm 1}}{\zeta^1\Pi_g} \\
        &\qquad + \sum_{\xi} \frac{\mel{\xi^3\Pi_{g, 0}}{H_\text{SO}}{\mathrm{b}^1\Sigma^+_g}^*}{E(\mathrm{b}^1\Sigma^+_g) - E(\xi^3\Pi_{g, 0})} \mel{\xi^3\Pi_{g, 0}}{\hat{L}_{\pm 1}}{\mathrm{X}^3\Sigma^-_{g, 1}}
    \end{split} \\
    \begin{split}
         &\quad = - \alpha_{\mathrm{X}, \mathrm{b}} \mel{\mathrm{X}^3\Sigma^-_{g, 0}}{\hat{S}_{\pm 1}}{\mathrm{X}^3\Sigma^-_{g, 1}} + \sum_{\zeta} \alpha_{\zeta, \mathrm{X}} \mel{\mathrm{b}^1\Sigma^+_g}{\hat{L}_{\pm 1}}{\zeta^1\Pi_g} \\
         &\qquad + \sum_{\xi} \alpha_{\xi, \mathrm{b}}^* \mel{\xi^3\Pi_{g, 1}}{\hat{L}_{\pm 1}}{\mathrm{X}^3\Sigma^-_{g, 1}} ,\label{eq:b-X_d1}
    \end{split}
\end{align}
which is composed of both spin and orbital angular momenta. However the relative strengths of the spin-orbit couplings imply that the primary contribution arises as a result of intensity borrowing from the spin-flip transitions between the $\Sigma = 0$ and $\Sigma = \pm 1$ sub-levels of the \Xstate\ state through spin-orbit coupling between the \bstate\ and \Xstate\ states. Electronic orbital magnetic moments contribute only weakly to this branch through spin-orbit mixing of the \Xstate\ and \sstates\ states, and of the \bstate\ and \tstates\ states.

Magnetic dipole transitions in the \astate\ -- \Xstate\ band are considerably weaker, and enabled only through spin-orbit coupling of the \astate\ and \Xstate\ states with the \tstates\ and \sstates\ states, respectively, as given by
\begin{align}
    \begin{split}
        &\mel{\mathrm{a}^1\Delta_g}{d_{\pm 1}}{\mathrm{X}^3\Sigma^-_{g,1}}_p \\
        &\quad= \sum_\xi \frac{\mel{\xi^3\Pi_{g, 2}}{H_\text{SO}}{\mathrm{a}^1\Delta_g}^*}{E(\mathrm{a}^1\Delta_g) - E(\xi^3\Pi_{g, 2})}
        \mel{\xi^3\Pi_{g, 2}}{\hat{L}_{\pm 1}}{\mathrm{X}^3\Sigma^-_{g, 1}} \\
        &\qquad + \sum_{\zeta} \frac{\mel{\zeta^1\Pi_g}{H_\text{SO}}{\mathrm{X}^3\Sigma^-_{g, 1}}}{E(\mathrm{X}^3\Sigma^-_{g, 1}) - E(\zeta^1\Pi_g)}
        \mel{\mathrm{a}^1\Delta_g}{\hat{L}_{\pm 1}}{\zeta^1\Pi_g}
    \end{split} \\
    &\quad= \sum_\xi \alpha_{\xi, a}^* \mel{\xi^3\Pi_{g, 2}}{\hat{L}_{\pm 1}}{\mathrm{X}^3\Sigma^-_{g,1}} + \sum_{\zeta} \alpha_{\zeta, X} \mel{a^1\Delta_g}{\hat{L}_{\pm 1}}{\zeta^1\Pi_g} . \label{eq:a-X_d1}
\end{align}

Finally, the \Xstate\ -- \Xstate\ magnetic dipole line strength, composed of rotation-vibration and pure rotational transitions arises as a result of two magnetic dipole moments. The first couples spin sublevels in the \Xstate\ state, and the second is composed of the diagonal spin magnetic moment, as given by
\begin{align}
    \mel{\mathrm{X}^3\Sigma^{-}_{g, 1}}{d_{\pm 1}}{\mathrm{X}^3\Sigma^{-}_{g, 0}}_p &= \mel{\mathrm{X}^3\Sigma^{-}_{g, 1}}{S_{\pm 1}}{\mathrm{X}^3\Sigma^{-}_{g, 0}} \label{eq:X-X_d1}\\
    \mel{\mathrm{X}^3\Sigma^{-}_{g, \Omega}}{d_{0}}{\mathrm{X}^3\Sigma^{-}_{g, \Omega}}_p &= \mel{\mathrm{X}^3\Sigma^{-}_{g, \Omega}}{\hat{S}_z}{\mathrm{X}^3\Sigma^{-}_{g, \Omega}}. \label{eq:X-X_d0}
\end{align}

\section{Spectroscopic Model}
\label{sec:spectroscopic_model}

\begin{figure}[h]
    \centering
    \includegraphics[width=.8\linewidth]{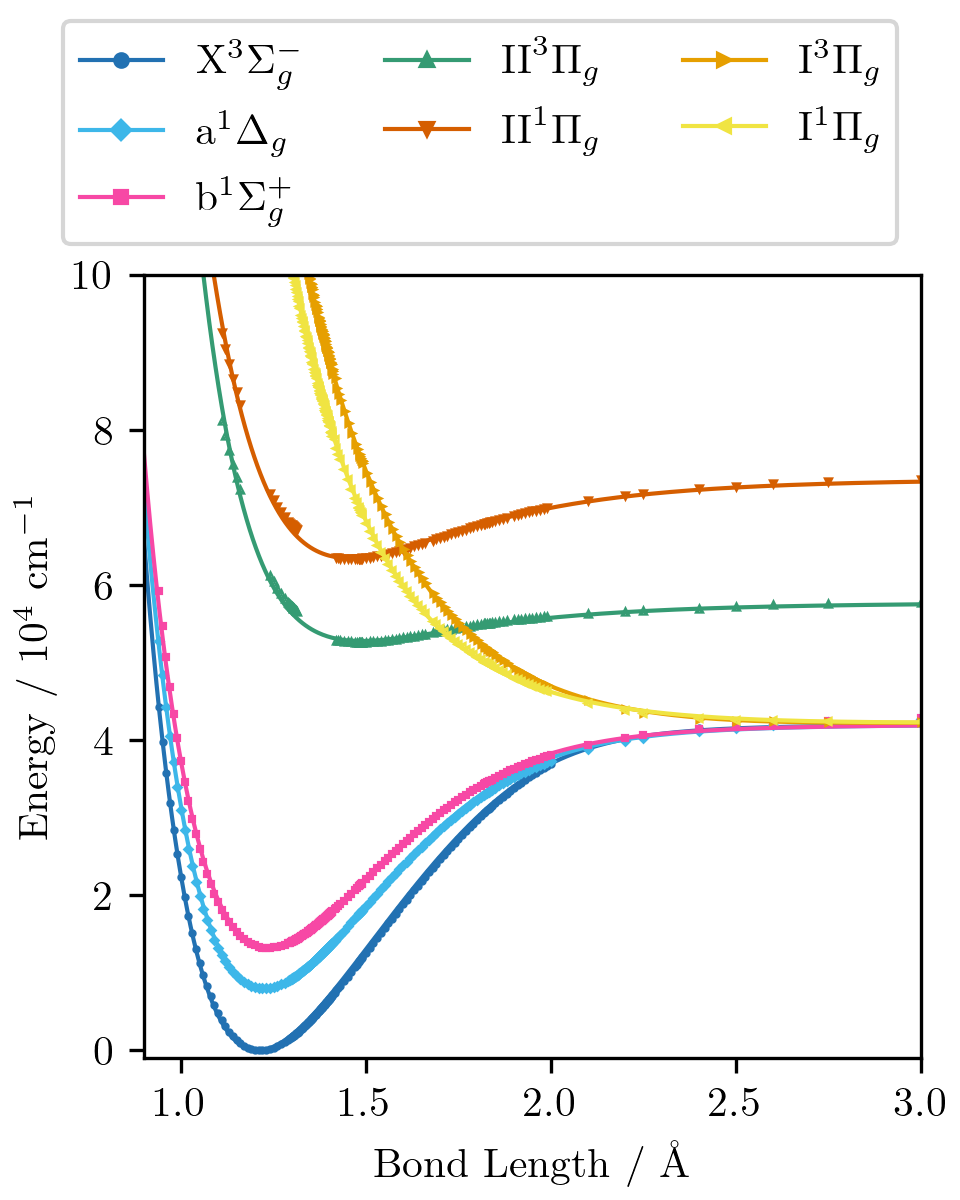}
    \caption{The PECs obtained from \ai\ electronic structure calculations (dots) along with the continuous curves (solid lines) obtained by fitting the analytic potential energy functions to these \ai\ data.}
    \label{fig:pecs}
\end{figure}

\begin{figure}[h]
    \centering
    \includegraphics[width=.8\linewidth]{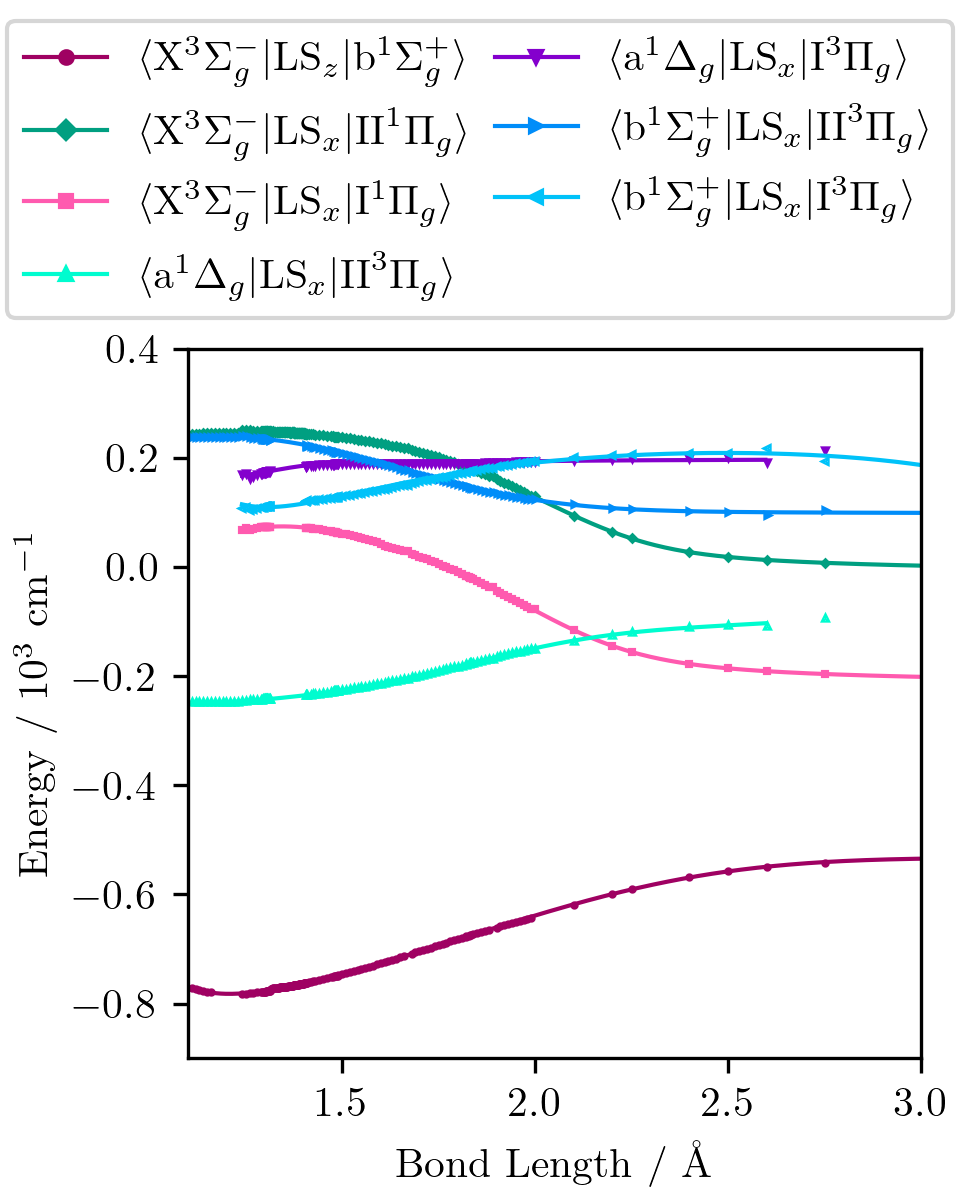}
    \caption{The SOCs obtained from \ai\ electronic structure calculations (dots) along with the continuous curves (solid lines) obtained by fitting the analytic coupling functions to these \ai\ data.}
    \label{fig:socs}
\end{figure}

\begin{figure}[h]
    \centering
    \includegraphics[width=.8\linewidth]{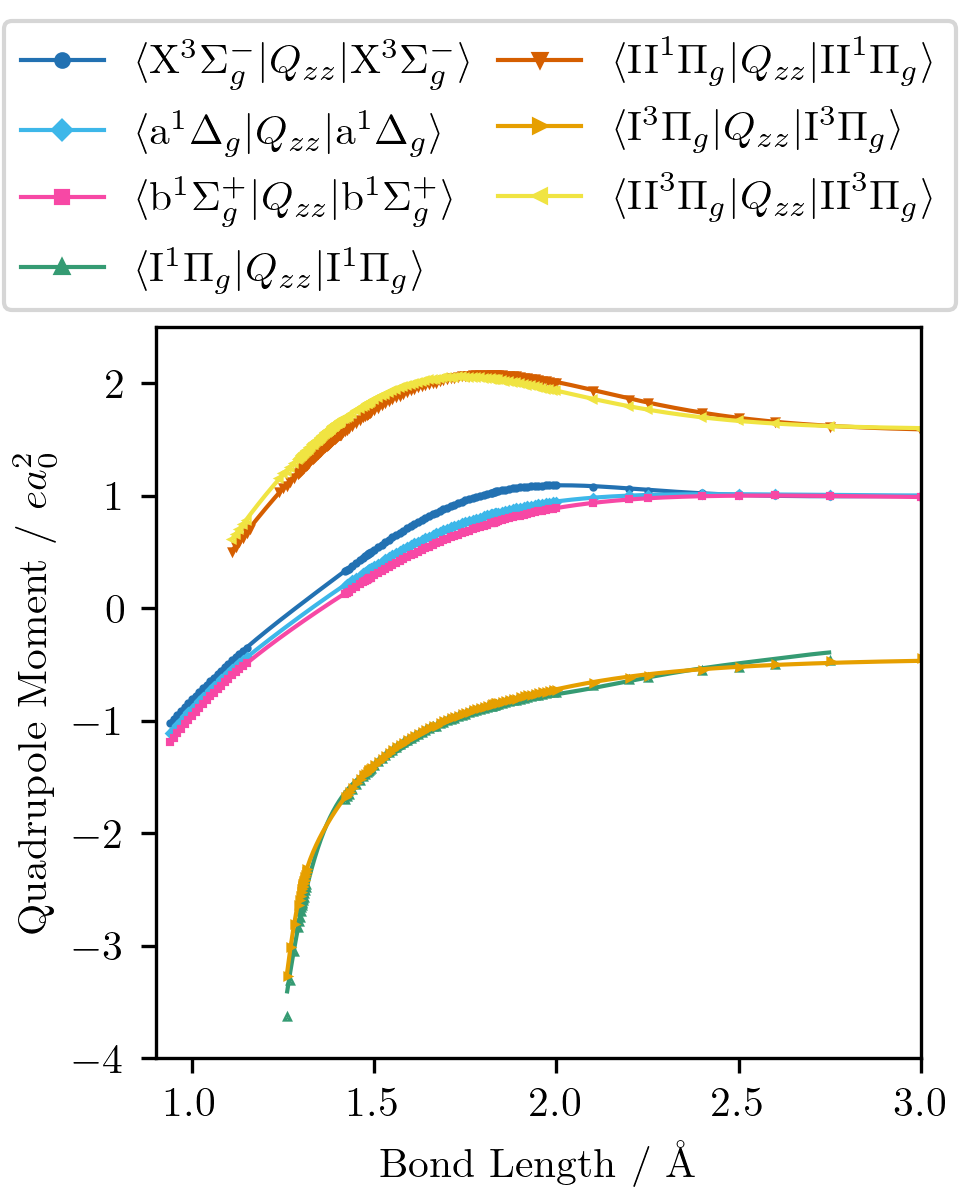}]
    \caption{The permanent EQMCs obtained from \ai\ electronic structure calculations (dots) along with the continuous curves (solid lines) obtained by fitting the analytic coupling functions to these \ai\ data.}
    \label{fig:eqmcs_1}
\end{figure}

\begin{figure}[h]
    \centering
    \includegraphics[width=.8\linewidth]{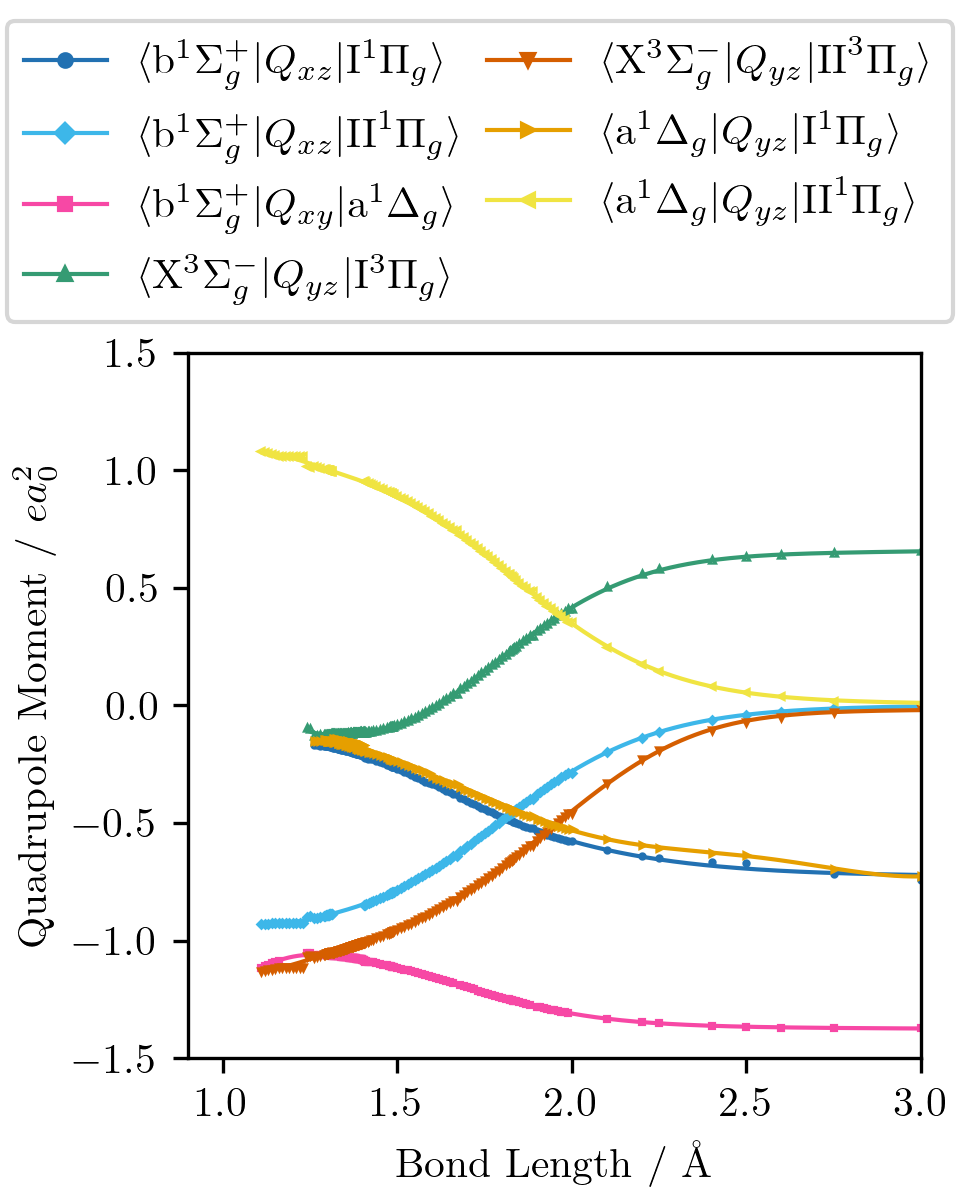}
    \caption{The transition EQMCs obtained from \ai\ electronic structure calculations (dots) along with the continuous curves (solid lines) obtained by fitting the analytic coupling functions to these \ai\ data.}
    \label{fig:eqmcs_2}
\end{figure}

\begin{figure}[h]
    \centering
    \includegraphics[width=.8\linewidth]{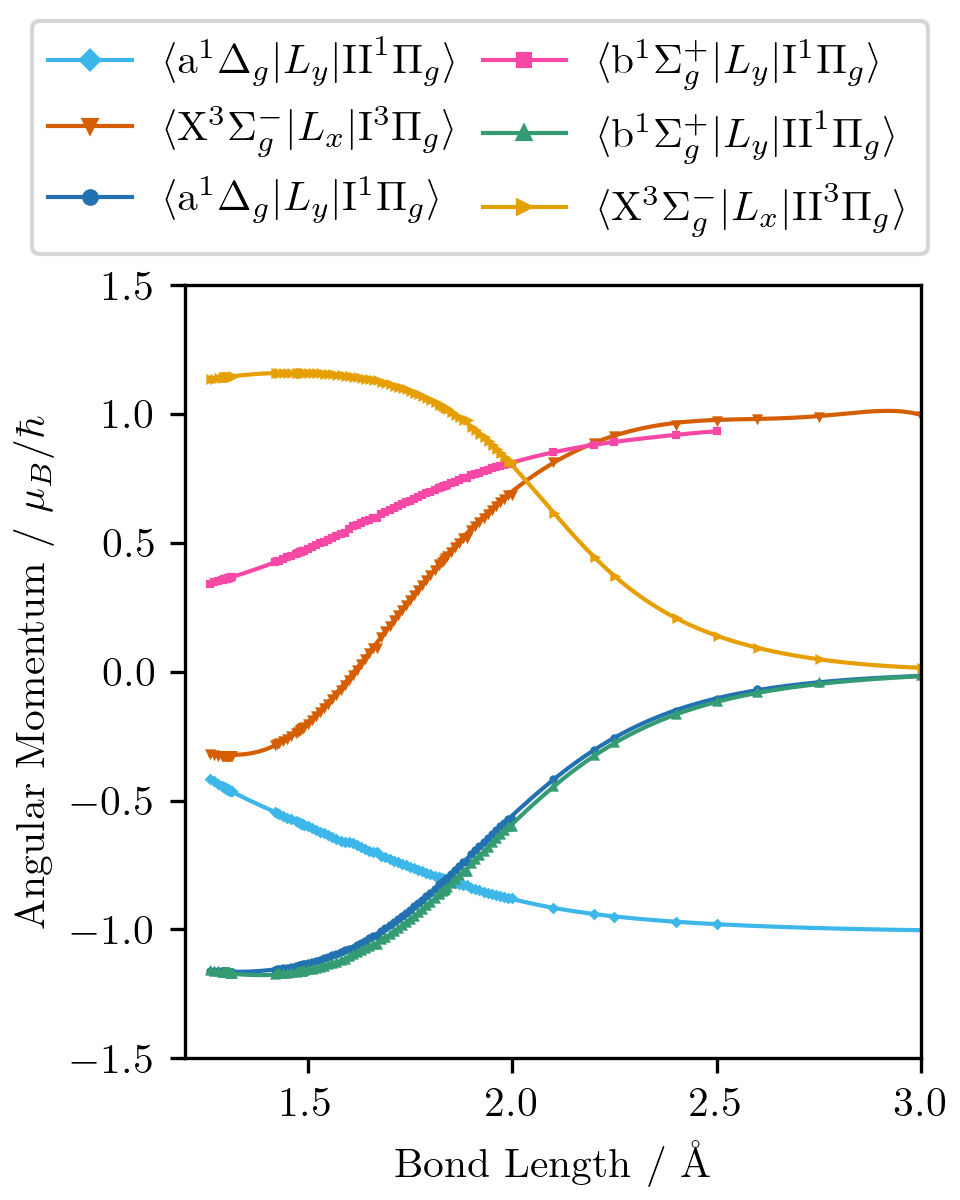}
    \caption{The EAMCs obtained from \ai\ electronic structure calculations (dots) along with the continuous curves (solid lines) obtained by fitting the analytic coupling functions to these \ai\ data.}
    \label{fig:eamcs}
\end{figure}

\subsection{Electronic Structure Calculations}
\label{sec:electronic_structure_calculations}

There are few \ai\ calculations of the electric quadrupole moment functions in the literature. The earliest calculations of the E2 moment for \oxygen\ were made by \citet{57KoMiKa.O2} in 1957, followed by \citet{74SaVaxx.O2} in 1974. In recent years accurate multipole moment calculations at room temperature were made by \citet{11BaCaHe.O2}, and \citet{14CoNtxx.O2}. In each case however, the quadrupole moment is given for only a single geometry. In 1997 \citet{97LaHaxx.O2} calculated the permanent E2 moment for a range of geometries. They use the aug-cc-pVQZ basis and MRCI method to obtain the permanent E2 moment of the $\Xstate$ state in the range \SIrange{2}{12}{\bohr}. \citet{04Mixxxx.O2} details MCSCF calculations for the permanent E2 moment for a narrow range of geometries between \SIrange{1}{1.8}{\angstrom}, obtaining a variety of values for $\mel{\mathrm{X}^3\Sigma^-_g}{Q_{zz}}{\mathrm{X}^3\Sigma^-_g}$ across different active space. The results broadly agree with experimental values and lie in the range \SIrange{-0.180}{-0.230}{\elementarycharge\bohr^2}. The need for further calculations of electric quadrupole moments in order to obtain a spectroscopic model of the \oxygen\ atmospheric bands is evident, particularly transition moments and the permanent moments for excited states.

The necessary set of electric quadrupole moment curves (EQMCs), spin-orbit coupling curves (SOCs), orbital electronic angular momentum curves (EAMCs), and potential energy curves (PECs), are obtained with electronic structure calculations using the \textsc{Molpro} program\citep{molpro}. Initially MCSCF calculations are performed with CAS(12,12) and an diffusion augmented, correlation-consistent, polarized quintuple-zeta basis set (aug-cc-pV5Z). The optimized orbitals from this MCSCF calculation are then used as the orbitals for a subsequent MRCI calculation in the same active space.

The MRCI calculations include a variety of symmetry groups and spin multiplicities in order to obtain a complete set of wavefunctions and potential energy curves for the required electronic states, and are detailed in Tab. \ref{tab:mrci_calculations}. Calculations are made for a number of geometries in the range \SIrange{0.090}{0.300}{\nano\meter} for the \Xstate, \astate\ and \bstate\ states, and in the range \SIrange{0.094}{0.300}{\nano\meter} for the $\mathrm{1}^1\Pi_g$, $\mathrm{2}^1\Pi_g$ $\mathrm{1}^3\Pi_g$, and $\mathrm{2}^3\Pi_g$ states.
For some geometries the MOLPRO calculations fail to converge and these are omitted from the final dataset.

\begin{table}[h]
    \centering
    \caption{This table details the number of states calculated for a given combination of irreducible symmetry group and spin multiplicity. It also shows the label of the corresponding states obtained from each calculation.}
    \label{tab:mrci_calculations}
    \begin{tabular}{cccc}
        \hline
        Symmetry & ($2S + 1$) & No. States & State(s) \\
        \hline
        $A_g$ & 1 & 2 & \bstate, \astate$(xx)$ \\
        $B_{1g}$ & 3 & 1 & \Xstate \\
        $B_{1g}$ & 1 & 1 & \astate$(xy)$ \\
        $B_{2g}$ & 1 & 2 & \Isstate$(x)$, \IIsstate$(x)$ \\
        $B_{3g}$ & 1 & 2 & \Isstate$(y)$, \IIsstate$(y)$ \\
        $B_{2g}$ & 3 & 2 & \Itstate$(x)$, \IItstate$(x)$ \\
        $B_{3g}$ & 3 & 2 & \Itstate$(y)$, \IItstate$(y)$ \\
        \hline
    \end{tabular}
\end{table}

The wavefunctions obtained are used in subsequent MRCI calculations of the seven spin-orbit couplings, nine quadrupole moments, and the six orbital angular momentum curves required to reproduce the matrix elements in Eqs. (\ref{eq:b-X_Q20}) - (\ref{eq:b-X_d1}). In addition to the quadrupole moments present in Eqs. (\ref{eq:b-X_Q20}) - (\ref{eq:a-X_Q21}), the permanent quadrupole moments of the \astate, and the four $\Pi$ states are also obtained. Finally, curves involving the $\Pi$ states are transformed to the diabatic representation using the numerical procedure described by \citet{22BrYuKi.SO}. These diabatic curves are presented in Figs. \ref{fig:pecs} - \ref{fig:eamcs}.

A challenge arises in evaluating the accuracy of \ai\ quadrupole moment curves due to a dearth of experimental data. Typically such measurements consist of only a single value, usually vibrationally averaged over the $v=0$ ground state. In 1968 \citet{68BuDiDu.O2} obtained a value $Q^{(2)}_0 = -0.30 \pm 0.08$ \si{\elementarycharge\bohr\squared} via pressure induced birefringence. \citet{77CoBixx.O2}, and \citet{76BiCoxx.O2} obtain $|Q^{(2)}_0| = 0.25$ \si{\elementarycharge\bohr\squared} via far-infrared spectra, which is in agreement with the measurement of  $|Q^{(2)}_0| = 0.22$ \si{\elementarycharge\bohr\squared} made by \citet{75Evxxxx.O2}. More recently \citet{14CoNtxx.O2} accurately measured the room-temperature quadrupole moment via electric-field-gradient-induced birefringence (EFGIB) to be $Q^{(2)}_0 = -0.2302 \pm 0.0061$ \si{\elementarycharge\bohr\squared}. The ground state vibrationally averaged quadrupole moment obtained via \Duo\ from the \ai\ data presented above has a value of $Q^{(2)}_0 = -0.223$ \si{\elementarycharge\bohr\squared}. The uncertainty quoted by Couling and Ntombela is the standard deviation, and so the \ai\ value obtained in the present work lies well within two standard deviations from their mean experimental value.

\subsection{Analytic Curves}
\label{sec:analytic_curves}

In order to fill in gaps in the \ai\ data where the MOLPRO calculations fail to converge, and also to eliminate discontinuities that are characteristic of \ai\ calculations across multiple geometries, we fit analytic functions for each of the curves described in Sec. \ref{sec:electronic_structure_calculations}. This also allows one to obtain spectroscopic parameters for the potential energy curves, and other characteristic quantities. The parameters obtained from fitting to the ab initio data can subsequently be used as a starting point for the empirical refinement of the calculated energy levels to experimental energies.

\subsubsection{Potential Energy Curves}

To begin we fit the Morse/long-range (MLR) potential energy function to the five bound states under consideration. The Morse/long-range function was introduced by \citet{06LeHuJa.fu} and later refined by \citet{09LeDaCo.fu}, and improves on the well-known Morse potential by accounting for the long-range behaviour of molecular potential energy surfaces. We fit a form of the MLR potential described by the following expression,
\begin{equation}
    V(r) = T_{\rm e} + (A_{\rm e} - T_{\rm e}) \left(1 - \frac{u(r)}{u(r_{\rm e})} e^{-\beta(r) y_p^{r_{\rm e}}(r)} \right)^2,
    \label{eq:mlr_potential}
\end{equation}
with $T_{\rm e}$ the potential minimum and $A_{\rm e}$ the dissociation energy, relative to the zero-point energy. The polynomial function $\beta(r)$ and the long-range function $u(r)$ ensure the function has the correct long-range behaviour,
\begin{gather}
    \beta(r) = y^{r_\text{ref}}_p(r)\beta_\infty \left( 1 - y_p^{r_\text{ref}}(r) \right) \sum_{i = 0}^{N_\beta} \beta_i (y^{r_\text{ref}}_{q})^i \label{eq:mlr_polynomial}, \\
    y_p^{r_x}(r) = \frac{r^p - r_x^p}{r^p + r_x^p}, \label{eq:surkus_variable}
\end{gather}
where $p$ is an integer greater than 1, the value of which is a hyper-parameter of the fitting procedure. The long-range function is $u(r) = \sum_n \frac{C_n}{r^n}$, where one or more of the coefficients $C_n$ may be equal to zero. In the limit $r \to \infty$, the function approaches the $\beta_\infty = \ln \left( \frac{2\mathcal{D}_{\rm e}}{u(r_{\rm e})} \right)$ where $r_{\rm e}$ is the equilibrium bond length, and $r_\text{ref}$ is some reference geometry. In all fits we select $r_\text{ref} \vcentcolon = r_{\rm e}$ for simplicity.

For the two dissociative potentials (\Isstate\ and \Itstate) we fit a repulsive potential in the form of a Laurent power series in the radial distance,
\begin{equation}
    V(r) = T_{\rm e} + \sum_i \frac{a_i}{r^i},
    \label{eq:dissociative_potential}
\end{equation}
where $a_i$ are the fitted coefficients. The fits for all potential energy curves are performed with the Python library \texttt{scipy}, using the Levenberg-Marquadt (LM) algorithm in the case of the dissociative potentials and the \textit{trust region reflective} (TRF) algorithm with a soft L1 loss function in the case of the bound potentials. For each of the bound potentials we set a single dissociation parameter, namely $C_6$, to be non-zero and fix the value as $C_6 = \num{2.95e5}$. The equilibrium bond length is bounded to remain within the range \qtyrange{0.9}{3.0}{\angstrom}, and the dissociation and excitation energies are bounded in the range of positive real numbers.

In Table \ref{tab:spectroscopic_parameters} we compare several key spectroscopic parameters obtained from the bound potentials obtained in this work to experimental and other \ai\ values.

\begin{table*}[h]
    \centering
    \caption{Key spectroscopic parameters obtained for the bound potentials presented in this work, compared to those obtained by \citet{14LiShSu.O2} and experimental parameters from \citet{04RuPiMo.O2} and \citet{79HuHexx.O2}. Values in square brackets indicate uncertain data, and those in rounded brackets indicate vibrationally averaged values obtained for the ground $v=0$ state.}
    \label{tab:spectroscopic_parameters}
    \renewcommand{\arraystretch}{.8}
    \begin{tabular}{rcccccc}
        \hline
         & $D_{\rm e}$ (\unit{\electronvolt}) & $T_{\rm e}$ (\unit{\per\cm}) & $R_{\rm e}$ (\unit{\nm}) & $\omega_{\rm e}$ (\unit{\per\cm}) & $B_{\rm e}$ (\unit{\per\cm}) & $10^2\alpha_{\rm e}$ (\unit{\per\cm}) \\
         \hline
         \Xstate \hspace{2em}     & 5.2146 & 0.0      & 0.12078 & 1590.16   & 1.4478 & 1.6207 \\
         Exp. \citep{04RuPiMo.O2} & 5.2142 & 0.0      & --      &--         &--      & --     \\
         Exp. \citep{79HuHexx.O2} & 5.2132 & 0.0      & 0.12075 & 1580.19   & 1.4456 & 1.59   \\
         Cal. \citep{14LiShSu.O2} & 5.2203 & 0.0      & 0.12068 & 1581.61   & 1.4376 & 1.2539 \\
         \astate \hspace{2em}     & 4.2313 & 7930.39  & 0.12229 & 1495.69   & 1.4083 & 1.4283 \\
         Exp. \citep{72Krxxxx.O2} & 4.2258 & 7918.11  & 0.12157 & (1509.3)  & 1.4263 & 1.71   \\
         Exp. \citep{79HuHexx.O2} &--      & 7918.1   & --      & [1483.5]  & 1.4264 & 1.71   \\
         Cal. \citep{14LiShSu.O2} & 4.2258 & 7776.43  & 0.12147 & 1491.07   & 1.3814 & 0.4238 \\
         \bstate \hspace{2em}     & 3.5786 & 13195.4 & 0.12366 & 1423.20   & 1.3798 & 1.6138 \\
         Exp. \citep{72Krxxxx.O2} & 3.5772 & 13195.31 & 0.12268 & (1432.67) & 1.4005 & 1.8169 \\
         Cal. \citep{14LiShSu.O2} & 3.6058 & 13099.92 & 0.12258 & 1438.65   & 1.4030 & 1.8018 \\
         \IItstate \hspace{2em}   & 0.6135 & 52800.3  & 0.14880 &  625.40   & 0.9520 & 4.1312 \\
         \IIsstate \hspace{2em}   & 1.2633 & 63500.6  & 0.14540 &  880.86   & 0.9964 & 2.0501 \\
         \hline
    \end{tabular}
\end{table*}

\subsubsection{Spin-Orbit Coupling Curves}

The use of a polynomial decay expansion as an analytic representation of the spin-orbit coupling interaction was introduced by \citet{17PrJaLo} and has been used in a number of diatomic spectroscopic models (see e.g \citet{22SeClYu} or \citet{22YuNoAz}). The polynomial decay function has the form
\begin{equation}
    F(r) = \sum_{i=0}^{N_B} B_i z^i \left(1 - y^{r_\text{ref}}_p\right) + y^{r_\text{ref}}_p B_\infty,
    \label{eq:polynomial_decay}
\end{equation}
where $y_p$ is the \v{S}urkus variable defined by equation (\ref{eq:surkus_variable}), and $z$ is taken as the damped coordinate
\begin{equation}
    z = (r - r_\text{ref}) e^{-\beta_2(r - r_\text{ref})^2 - \beta_4(r - r_\text{ref})^4}.
    \label{eq:damped_coordinate}
\end{equation}

We use the same representation to parameterize all seven SOCs in the present work. Initially we attempt to fit a 3rd order polynomial for all SOCs, using TRF and a linear loss function. For each SOC the order of the polynomial is then increased until the fit converges with an $R^2$ value greater than 0.95. This ensures a good fit whilst minimizing the degree of overfitting.

\subsubsection{Transition Moment Curves}

Analytic representations of the EQMCs have not been widely used in the existing literature. We find that the LM algorithm is successful in parameterizing the fourteen quadrupole moment curves with the same polynomial decay function as for the spin-orbit curves. Though we find that higher order polynomials are often required in order to reach convergence. To obtain good fits of the EAMCs we find a variety of analytic forms are needed. The polynomial decay function defined by Eqs. (\ref{eq:surkus_variable}) and (\ref{eq:damped_coordinate}) is used to represent the \Mel{\astate}{L_y}{\IIsstate} curve, with a 6-th order polynomial. The \Mel{\Xstate}{L_x}{\Itstate} EAMC is represented by a simple polynomial expansion of the 10-th order with the form
\begin{equation}
    F(r) = T_{\rm e} + a_1 (r - r_\text{ref}) + a_2 (r - r_\text{ref})^2 + \cdots
    \label{eq:polynomial}
\end{equation}

For all other EAMCs we find a good fit is possible using the so-called irregular Chebyshev polynomial, which was originally introduced by \citet{22MeUsxx.fu} to represent electronic dipole moment curves and has the form
\begin{equation}
    F(r) = \frac{
        \left(1 - e^{c_2r}\right)^3
    }{
        \sqrt{ \left(r^2 - c_3^2\right)^2 + c_4^2 } \sqrt{ \left(r^2 - c_5^2\right)^2 + c_6^2}
    }
    \sum_{k=0}^6 b_k T_k\left(z_1(r)\right),
    \label{eq:irregular_chebyshev}
\end{equation}
where $T_k(z)$ are the Chebyshev polynomials of the first kind, $b_k$ are the expansion coefficients and
\begin{equation}
    z_1(r) = 1 - 2e^{-c_1 r}
    \label{eq:chebyshev_variable}
\end{equation}
maps the $r \in [0, \infty]$ half-infinite interval to the $z \in [-1, +1]$ finite interval. The parameters $b_i$ and $c_i$ are the fitting parameters. As with the SOCs, we find a variety of polynomial orders are required to fit individual EAMCs. The parameters for all fitted curves are given in the Supplementary Information\dag.

\subsubsection{Spin-spin Splitting Curve}

In addition to the \ai\ curves given above, we must also account for the spin splitting of the triplet ground state, which is crucial for the production of the \bstate\ -- \Xstate\ magnetic dipole transitions Eq. (\ref{eq:b-X_d1}). \citet{55TiStxa.O2} estimated the magnitude of the separation as \SI{1.17}{\per\cm}. The spin splitting of the \Xstate\ electronic state can be accounted for by including the phenomenological spin-spin operator term in the rovibronic Hamiltonian \cite{Duo}. \citet{02VaLoMi.O2} present calculations of the spin-orbit and spin-spin contributions to the zero-field splitting (ZFS) of the \Xstate\ ground state and estimate that the spin-spin contribution at the equlibrium geometry is approximately 40\% of the total energy difference between the $\Omega = 0$ and $\Omega = 1$ states, $D_{\textrm{SS}}$ = \SI{1.57}{\per\cm}. We provide the values presented in their work at different internuclear geometries to the \Duo\ program, which then performs a cubic spline interpolation on the grid of internuclear geometries described in Sec. \ref{sec:results}.

\begin{figure*}[p]
    \centering
    \includegraphics[width=.45\linewidth]{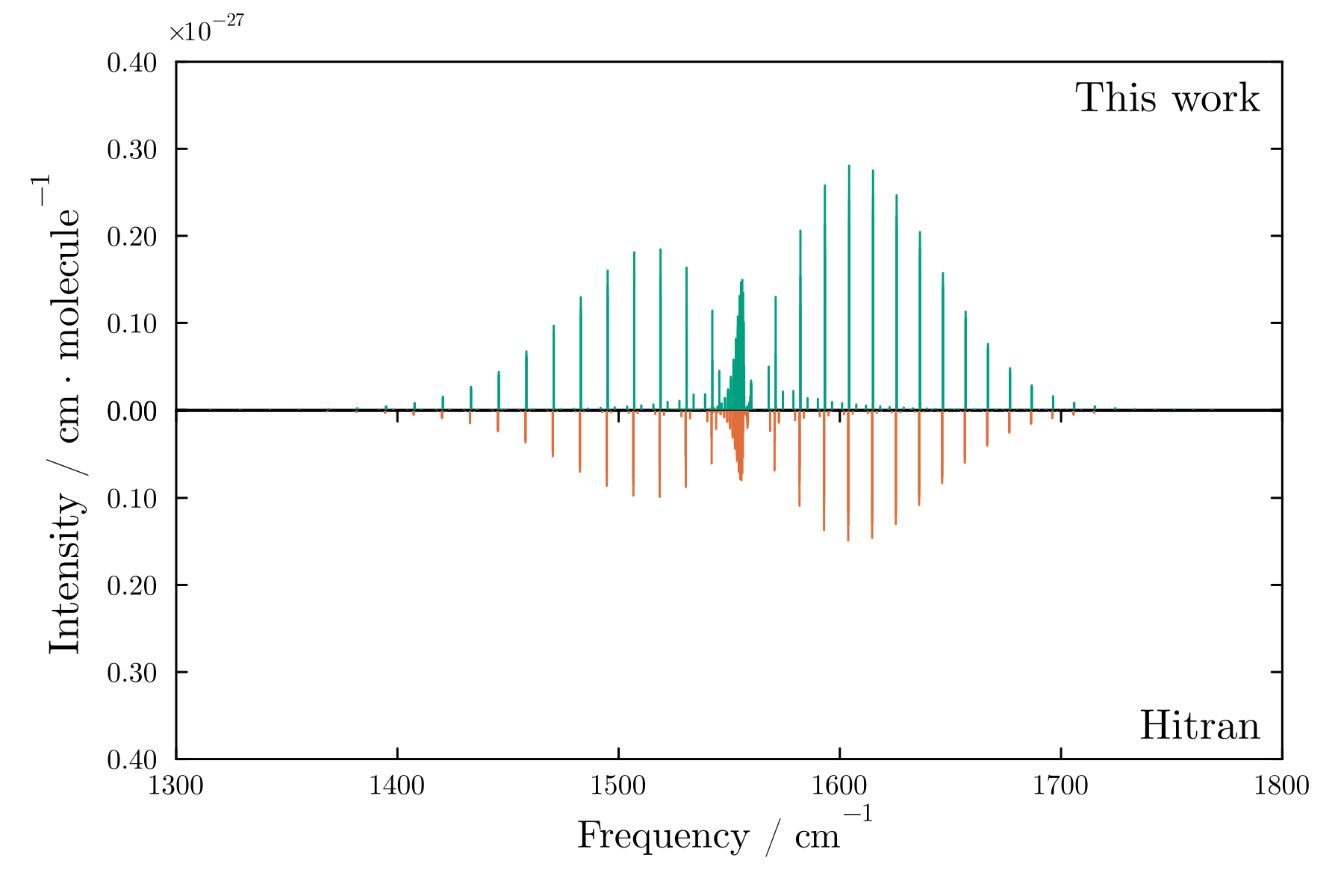}
    \includegraphics[width=.45\linewidth]{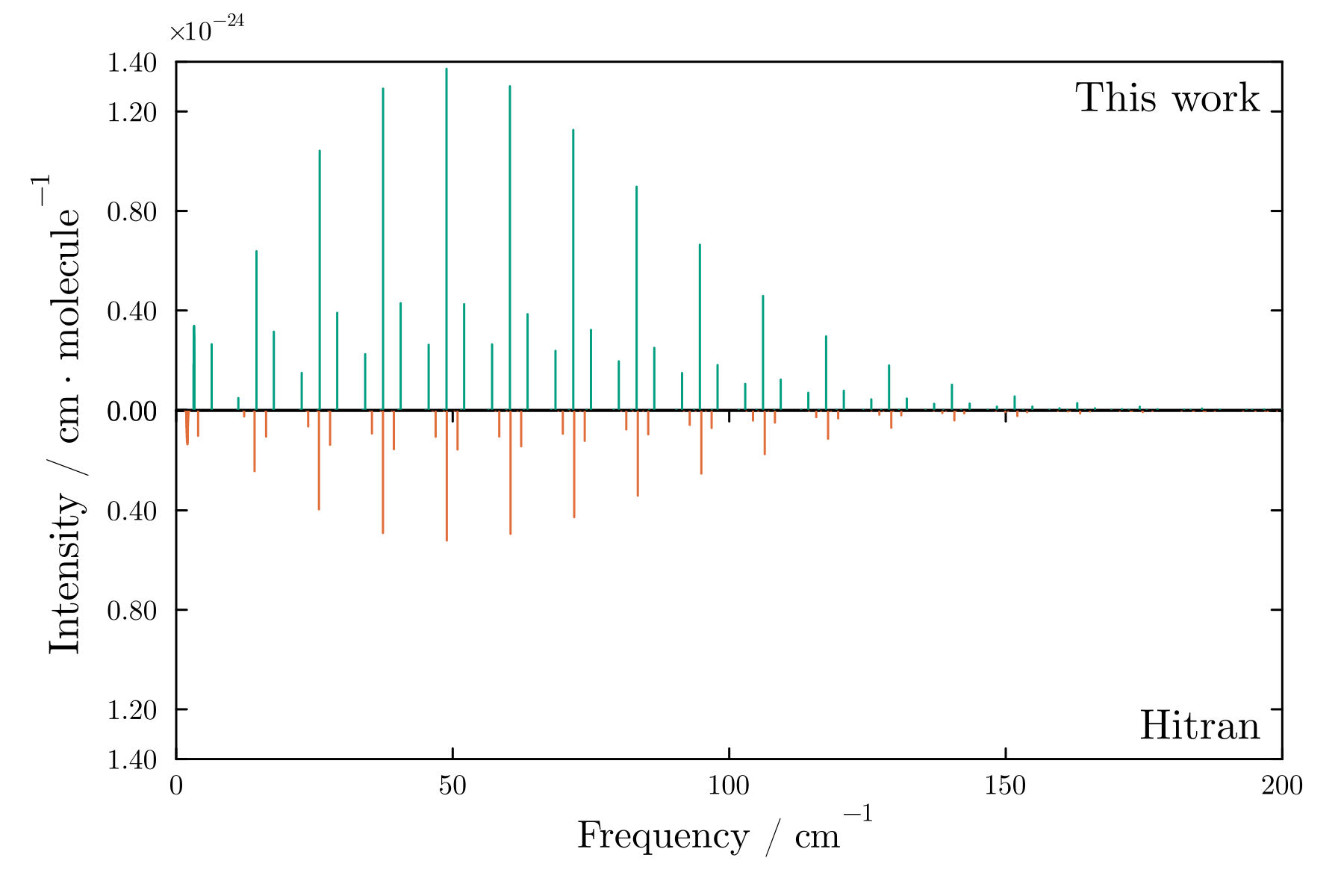}
    \includegraphics[width=.45\linewidth]{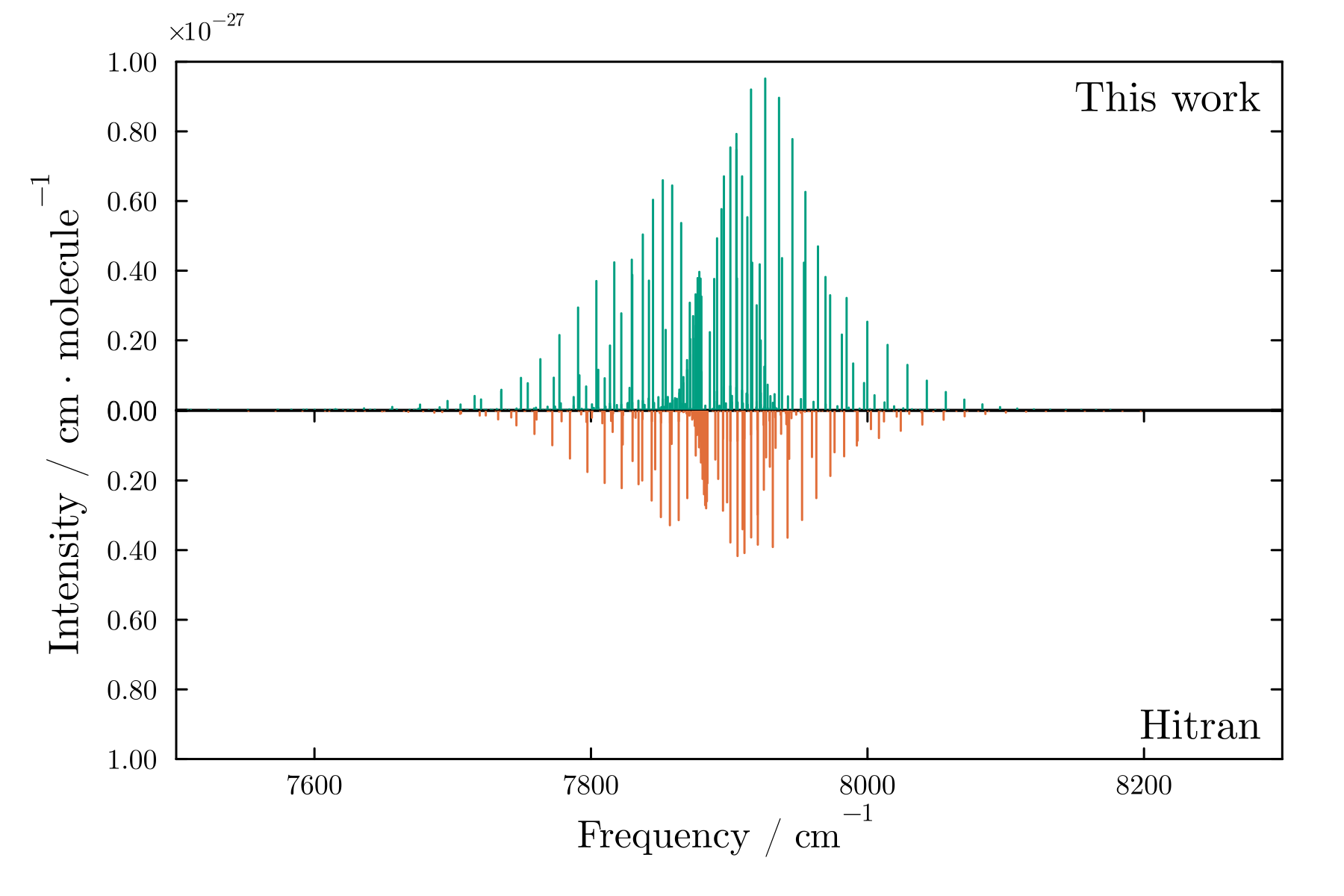}
    \includegraphics[width=.45\linewidth]{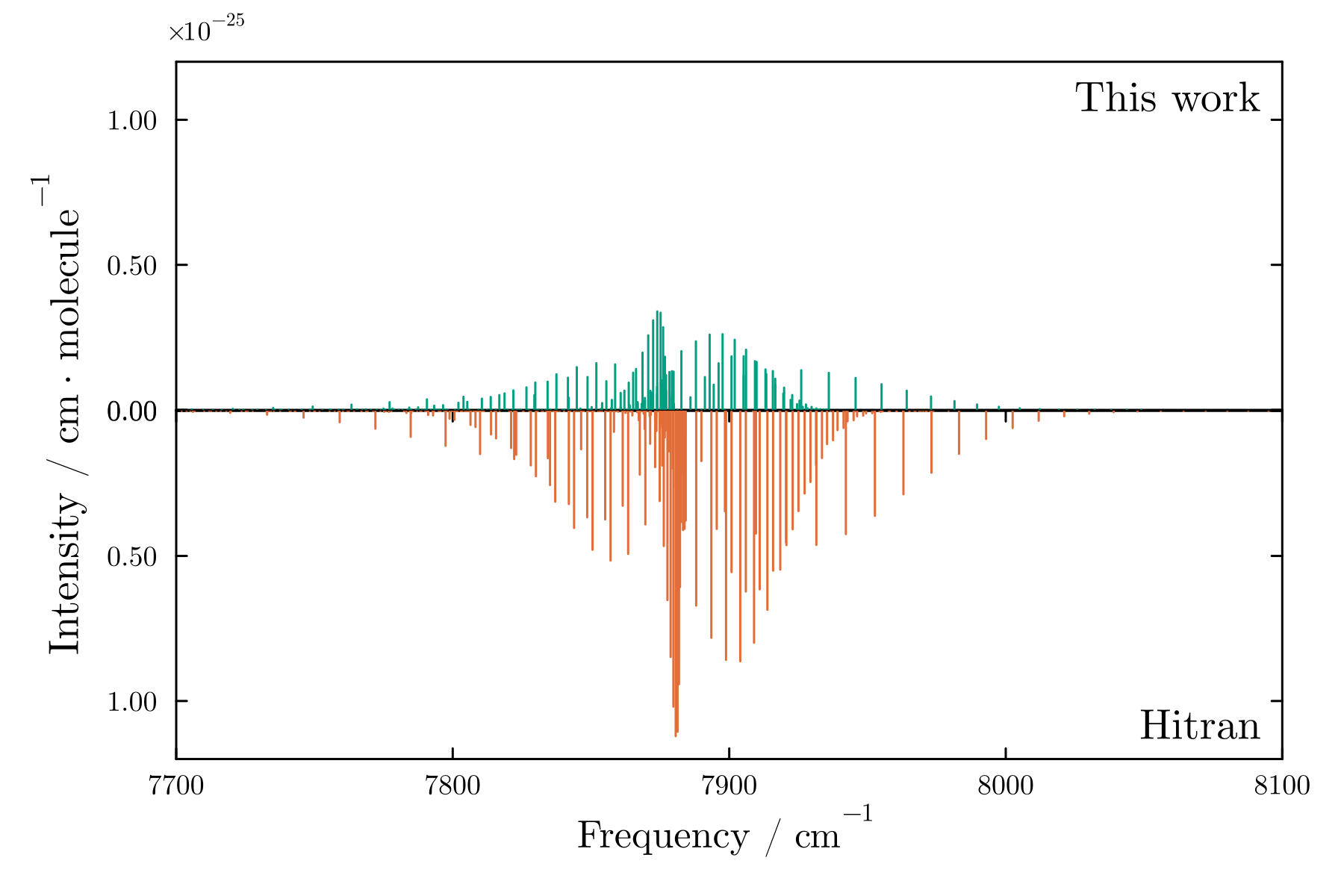}
    \includegraphics[width=.45\linewidth]{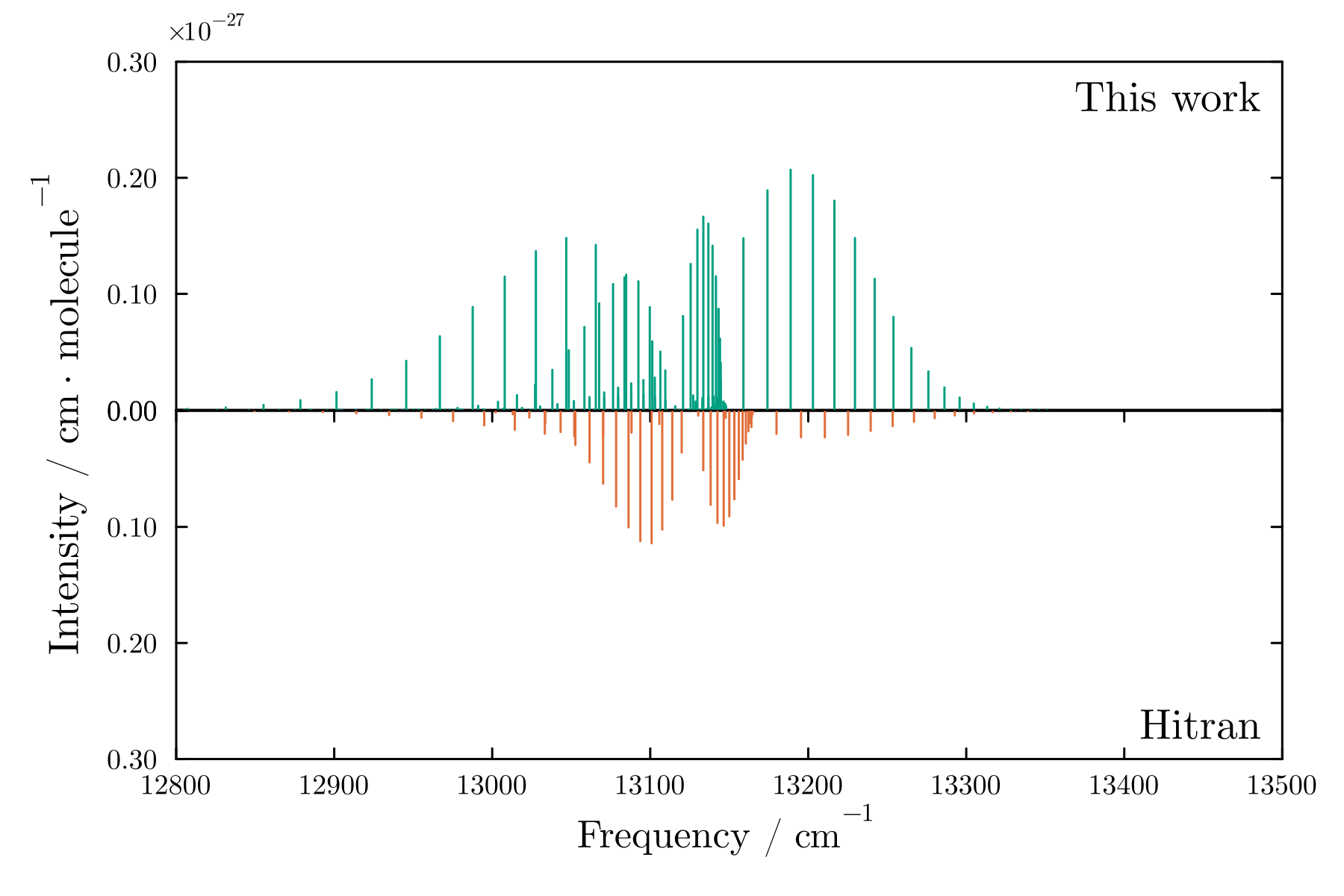}
    \includegraphics[width=.45\linewidth]{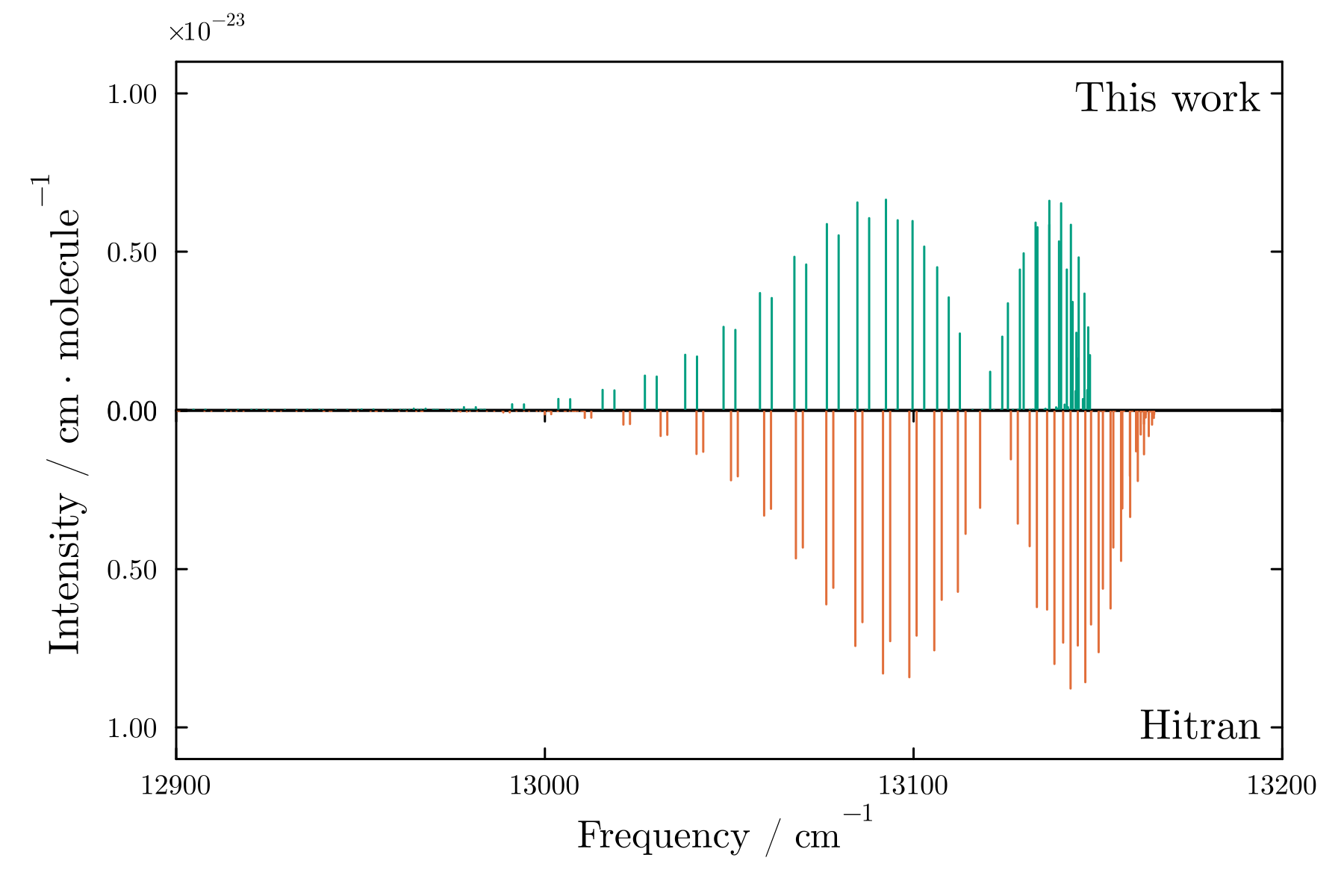}
    \caption{Comparison of the most intense electric quadrupole (left) and magnetic dipole (right) transitions in the \Xstate\ -- \Xstate\ (top), \astate\ -- \Xstate\ (middle) and \bstate\ -- \Xstate\ (bottom) bands. The pink dots indicate the transition intensities recorded in the HITRAN database. In each case we show the $v = 1 \leftarrow 0$ transitions except for the \Xstate\ -- \Xstate\ magnetic dipole band (top right), which is formed from the pure rotational $v = 0 \leftarrow 0$ transitions. The \Xstate\ -- \Xstate\ bands contain closely separated lines as a result of the spin splitting of the triplet state produces $\Omega$ sub-levels with only small energy differences. The closely separated group of transitions below \SI{10}{\per\cm} in the magnetic dipole moment \Xstate\ -- \Xstate\ $v = 0 \leftarrow 0$ plot are the fine structure transitions between spin sub-levels in the same total rotational ($J$) state.}
    \label{fig:abinitio_line_list_bands}
\end{figure*}

\section{Results}
\label{sec:results}

Using the analytic representation of the \ai\ curves obtained in Sec. \ref{sec:analytic_curves}, we build a spectroscopic model for the electric quadrupole and magnetic dipole transitions of \oxygen\ in the infrared and visible region of the electromagnetic spectrum. We include transitions between all states lower in energy than \SI{80000}{\per\cm}. In solving the nuclear Schr\"{o}dinger equation, a vibrational sinc-DVR basis set for each electronic state is defined on a grid of 1001 points in the range \SIrange{0.90}{3.00}{\AA}. The vibronic basis sets are then truncated to the lowest 30 vibrational levels in the case of the \Xstate, \astate, and \bstate\ states, and to the lowest 300 vibrational levels for the \IIsstate, \IItstate, \Isstate, and \Itstate states. We then solve for rotational levels up to $J = 50$. A large number of vibrational basis states are retained for the weakly bound and dissociative $\Pi$ state in order to represent continuum states above the dissociation energy \citep{22PeTeYu, 23YuWoHa}.

In this work we present a line list using a purely \ai\ model, without empirical refinement of the potential energy or coupling curves. In Fig. \ref{fig:abinitio_line_list} we show the absorption intensities for the electric quadrupole and magnetic dipole moments, respectively, above a threshold of \SI{e-30}{\cm\per\molecule}. In Fig. \ref{fig:abinitio_line_list_bands} we present stick spectra to compare the calculated intensities and line positions to accurate known data from the HITRAN database \cite{21GoRoHa}, namely the $\Xstate(v=1) \rightarrow \Xstate(v=0)$ (E2), $\Xstate(v=0) \rightarrow \Xstate(v=0)$ (M1), $\astate(v=0) \rightarrow \Xstate(v=0)$, and $\bstate(v=0) \rightarrow \Xstate(v=0)$ transitions. In addition to the six bands vibrational bands present in the HITRAN database, we also obtain intensities for 14 additional vibrational bands above the threshold intensity of \SI{e-30}{\cm\per\molecule}.

\begin{figure}
    \centering
    \includegraphics[width=\linewidth]{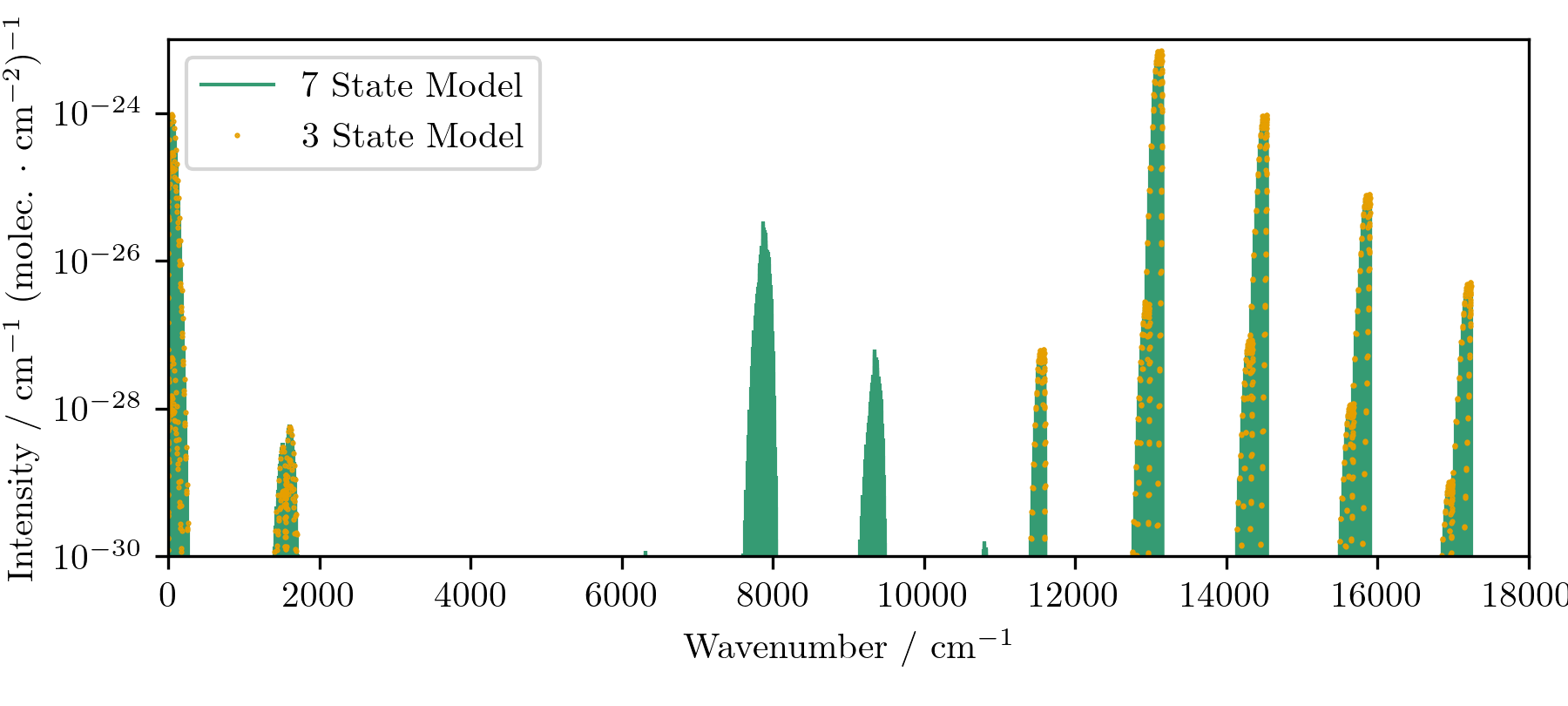}
    \caption{A comparison of the magnetic dipole line list obtained using the only the three low-lying electronic states to that obtained when the excited $\Pi$ states are included, which demonstrates the importance of the $\Pi$ states in producing intensities for the \astate\ -- \Xstate\ transitions.}
    \label{fig:abinitio_line_list_md3}
\end{figure}

The \ai\ model reproduces the expected intensities to the correct order of magnitude, with the largest discrepancies observed for the \textsuperscript{N}O and \textsuperscript{T}S branches of the \bstate\ -- \Xstate\ electric quadrupole transitions. These branches correspond to $\Delta N = -3$, $\Delta J = -2$ and $\Delta N = +3$, $\Delta J = +2$ transitions, respectively. Crucially, the combination of PECs and couplings present in the model successfully reproduces all of the observed rotational bands and the fine structure magnetic dipole transitions between spin sub-levels in the \Xstate. To illustrate the role of the excited $\Pi$ states we also reduce the model to a simple three state model that contains only the \Xstate, \astate, and \bstate\ with their respective couplings. Without the inclusion of the four $\Pi$ states, we find that no intensities for the \astate -- \Xstate\ magnetic dipole transitions are produced, and the \textsuperscript{P}O and \textsuperscript{R}S branches ($\Delta N = -1$, $\Delta J = -2$ and $\Delta N = +1$, $\Delta J = +2$) of the \bstate\ -- \Xstate\ electric quadrupole transitions are also not reproduced. In Fig. \ref{fig:abinitio_line_list_md3} we compare the intensities obtained with and without the inclusion of the $\Pi$ states.

\section{Conclusions}
\label{sec:conclusions}

The key result of this work is a novel set of \ai\ electronic structure calculations for various properties of 7 electronic states of \oxygen, namely the PECs of 7 electronic states, \Xstate, \astate, \bstate, \Isstate, \Itstate, \IIsstate\ and \IItstate, along with a complete set of corresponding SOCs, EAMCs and EQMCs. These \ai\ calculations have been used to produce a self-consistent \ai\ spectroscopic model describing the magnetic dipole and electric quadrupole spectra of the three main atmospheric bands of O$_2$, \bstate\ -- \Xstate\, \astate\ -- \Xstate and \Xstate\ -- \Xstate, which reproduces all the experimentally measured rotational branches within the dipole-forbidden electronic bands. We have shown that the highly excited $\Pi$ states, although weakly coupled, are crucial for properly describing the observed bands in the infra-red region. The \ai\ line list produced provides approximate intensities and line positions for the overtone transitions and rotational levels up to $J = 50$.

Due to the challenging nature of the electronic structure calculations, and the complex combination of couplings that contributes to each band, the line position and intensities of the purely \ai\ model do no accurately reproduce experimentally measured lines. However, it is expected that the line positions and intensities can be improved significantly with respect to experimental data by empirical refinement of the potential energy and coupling curves. In future work this can be achieved by optimising the parameters of the analytic functions outlined in Sec. \ref{sec:analytic_curves} in order to obtain an optimal fit to state energies, rather than to the \ai\ values of the potential energy and coupling moments. We propose that the MARVEL analysis of \citet{19FuHoKo.O2} is an ideal source of data with which to perform this refinement.

In this work we have restricted the model to consider only the first-order spin-orbit coupling perturbations and additional perturbations from other states such as the $1^5\Pi_g$ state, which has previously been considered in the context of oxygen airglow by \citet{20MiPaxx.O2}, have not been considered. Although these states are more highly excited and so coupled weakly around the equilibrium bond length, and longer bond lengths the strength of the coupling increases and these perturbations may provide small corrections to the rotational structure that further improve an empirically refined model.

\section*{Conflicts of interest}

There are no conflicts to declare.

\section*{Data availability}

The \ai\ data and the parameters of the analytic curves fitted to this data are provided as part of the ESI\dag. In addition to this data, we also provide the spectroscopic model presented in Sec.~\ref{sec:spectroscopic_model} in the form of an input file for the \Duo\ program, which itself is available online via the ExoMol GitHub (https://github.com/exomol/duo).

\section*{Acknowledgements}
This work was supported by the STFC Projects No. ST/Y001508/1 and ST/S506497/1. The authors acknowledge the use of the UCL Legion High Performance Computing Facility (Myriad@UCL) and associated support services in the completion of this work along with the Cambridge Service for Data Driven Discovery (CSD3), part of which is operated by the University of Cambridge Research Computing on behalf of the STFC DiRAC HPC Facility (www.dirac.ac.uk). The DiRAC component of CSD3 was funded by BEIS capital funding via STFC capital grants ST/P002307/1 and ST/R002452/1 and STFC operations grant ST/R00689X/1. DiRAC is part of the National e-Infrastructure.

This work was also supported by the European Research Council (ERC) under the European Union Horizon 2020 research and innovation programme through Advance Grant number 883830. This work is partly supported by the European Metrology Partnership project ``22IEM03 PriSpecTemp'' (Funder ID 10.13039/100019599), which received funding from EPM Programme co-financed by the Participating States and from the European Union's Horizon 2020 research and motivation program and participating states.

\providecommand*{\mcitethebibliography}{\thebibliography}
\csname @ifundefined\endcsname{endmcitethebibliography}
{\let\endmcitethebibliography\endthebibliography}{}

\end{document}